\documentclass[aps,prd,a4paper,superscriptaddress,nofootinbib,10pt, two column,groupedaddress]{revtex4-1}
\usepackage{amsmath,amssymb,mathtools,graphicx,multirow,dcolumn,bm,latexsym,soul,nicefrac}
\usepackage[dvipsnames]{xcolor}
\usepackage[colorlinks,linkcolor=blue,citecolor=blue,urlcolor=blue]{hyperref}
\newcounter{RomanNumber}
\usepackage[normalem]{ulem}

\allowdisplaybreaks
\newcommand{\be}{\begin{equation}}
\newcommand{\ee}{\end{equation}}
\newcommand{\bea}{\begin{eqnarray}}
\newcommand{\eea}{\end{eqnarray}}

\def\simlt{\mathrel{\lower0.6ex\hbox{$\buildrel {\textstyle <}
 \over {\scriptstyle \sim}$}}}

\def\bea{\begin{eqnarray}}
\def\eea{\end{eqnarray}}
\def\be{\begin{equation}}
\def\ee{\end{equation}}
\def\bes{\begin{split}}
\def\ees{\end{split}}
\def\ba{\begin{eqnarray}}
\def\ea{\end{eqnarray}}
\def\p{\partial}

\def\dd{{\rm d}}

\def\dr{{\rm d}r}

\def\m{{\rm m}}
\def\({\Big(}
\def\){\Big)}

\def\a{\alpha}
\def\b{\beta}

\def\m{\mu}
\def\n{\nu}

\def\th{\theta}

\def\l{\lambda}

\newcommand{\eg}{\textit{e.g.}}

\newcommand{\ii}{\mathrm{i}}
\newcommand{\eex}{\mathrm{e}}

\newcommand{\Ein}{\mathcal{G}}

\newcommand{\SE}{\mathcal{T}}

\newcommand{\const}{\mathsf{const}}

\newcommand{\sst}{\sin^2\!\th}

\newcommand{\aaee}{\text{\ae}}

\newcommand{\Ho}{\mathcal{H}}

\newcommand{\al}{\alpha}
\newcommand{\ga}{\gamma}

\newcommand{\de}{\delta}

\renewcommand\th{\theta}

\newcommand{\La}{\Lambda}

\newcommand{\om}{\omega}

\newcommand{\K}{\mathcal{K}}

\newcommand{\dl}{\partial}
\newcommand{\Dl}{\nabla}

\begin{document}

\title{The relation between general relativity and a class of Ho\v rava gravity theories}

\author{Nicola Franchini}
\author{Mario Herrero-Valea} 
\author{Enrico Barausse}
\affiliation{SISSA, Via Bonomea 265, 34136 Trieste, Italy and INFN Sezione di Trieste}
\affiliation{IFPU - Institute for Fundamental Physics of the Universe, Via Beirut 2, 34014 Trieste, Italy}

\begin{abstract}
Violations of Lorentz (and specifically boost) invariance
can make gravity renormalizable in the ultraviolet, as initially noted by Ho\v rava,
 but are increasingly constrained
in the infrared. At low energies, Ho\v rava gravity is
characterized by three dimensionless couplings,  $\alpha$, $\beta$ and $\lambda$, 
which vanish in the general relativistic limit. 
Solar system and gravitational wave experiments bound two of these couplings
($\alpha$ and $\beta$) to tiny values, 
but the third remains relatively unconstrained ($0\leq\lambda\lesssim 0.01-0.1$).
Moreover, demanding that (slowly moving) black-hole solutions are regular away from the central singularity requires $\alpha$ and $\beta$
to vanish {\it exactly}. 
Although a canonical constraint analysis
shows that the class of khronometric theories
resulting from these constraints ($\alpha=\beta=0$ and $\lambda\neq0$) 
cannot be equivalent to general relativity,
 even in vacuum, previous calculations of the dynamics of the
 solar system, binary pulsars and gravitational-wave generation
 show perfect agreement with general relativity. Here, we 
 analyze spherical collapse and compute 
 black-hole quasinormal modes, and find again that they
  behave {\it exactly} as in general relativity, as far as {\it observational}
  predictions are concerned. Nevertheless, we find that 
  spherical collapse leads to the formation of a regular {\it universal} horizon, 
  i.e. a causal boundary for signals of arbitrary propagation speeds, inside the
  usual event horizon for matter and tensor gravitons.
  Our analysis also confirms that  the additional scalar degree of freedom present alongside the spin-2 graviton of general relativity remains strongly coupled at low energies, even on curved backgrounds.
    These puzzling results suggest that any further bounds on Ho\v rava gravity will probably come from cosmology.
\end{abstract}

\maketitle

\section{Introduction}
Lorentz symmetry is one of the cornerstone of our understanding of theoretical physics, and has been tested to
exquisite precision in particle physics experiments~\cite{Kostelecky:2003fs,Kostelecky:2008ts,Mattingly:2005re,Jacobson:2005bg}. Bounds on Lorentz violations (LVs) in gravity are however 
much weaker~\cite{Jacobson:2008aj,Liberati:2013xla,Kostelecky:2010ze}. This is particularly interesting because violations of boost symmetry in gravity
may allow for constructing a theory of quantum gravity that is power counting (or even perturbatively) renormalizable
in the ultraviolet~\cite{Horava:2009uw,Barvinsky:2015kil}. This proposal, initially put forward by Ho\v rava~\cite{Horava:2009uw}, may still pass
particle physics tests of Lorentz symmetry if a mechanism is included to prevent ``percolation''
of large LVs from gravity to matter. Among such putative mechanisms are
renormalization group flows (whereby  Lorentz invariance may be recovered, at least in matter, in the infrared)~\cite{Chadha:1982qq,Bednik:2013nxa,Barvinsky:2017kob,Barvinsky:2019rwn},
accidental symmetries allowing for different degrees of LVs in gravity and matter~\cite{GrootNibbelink:2004za}, or the
suppression of LVs in matter via a large energy scale~\cite{Pospelov:2010mp}.

The infrared limit of Ho\v rava gravity, also known as khronometric theory, is characterized
by three dimensionless coupling parameters $\alpha$, $\beta$ and $\lambda$, in terms of which the theory's action is~\cite{Horava:2009uw,Blas:2009qj,Blas:2010hb}
\be
\begin{split}\label{eq:actionHL}
S=&\frac{1-\beta}{16\pi G}\int \dd T\dd^3x\, N\sqrt{\gamma} \Big(\K_{ij}\,\K^{ij}-\frac{1+\lambda}{1-\beta}\K^2\\
&+\frac{1}{1-\beta}{}^{(3)}{R}+\frac{\alpha}{1-\beta}a_i\,a^i\Big)+S_{\rm matter}[g_{\mu\nu},\Psi]\,,
\end{split}
\ee
in units where $c=1$ (used throughout this article), and where the
bare gravitational constant $G$ is related to the one measured on Earth and
in the solar system ($G_N$) by
\begin{equation}
    G_N=\frac{G}{1-\alpha/2}\,.
\end{equation}
The action is written in terms of a {\it preferred} spacetime foliation described by $T =\const$, and the metric has been decomposed in the 3+1 form\footnote{From now on, Latin indices will run only over space directions, while Greek indices will also include time.}
\begin{equation}
\dd s^2 = N^2\dd T^2 - \ga_{ij}\left(\dd x^i + N^i\dd T\right)\left(\dd x^j + N^j \dd T\right),
\end{equation}
where we recognise a lapse function $N$, a shift three-vector $N_i$ and the spatial three-metric $\ga_{ij}$. Also defined in terms of this decomposition are the other quantities appearing in the action, \textit{e.g.}~the determinant of the three-metric $\ga$; the extrinsic curvature of the foliation,
\begin{equation}
    \K_{ij} = -\frac{1}{2N}\left(\dl_T \ga_{ij} - D_i N_j - D_j N_i\right)\,,
\end{equation}
where the covariant derivative $D_i$ is defined with respect to $\ga_{ij}$; the three-dimensional Ricci scalar ${}^{(3)}{R}$; $\K=\K^{ij}\gamma_{ij}$; and $a_i=\p_i\ln N$. With $\Psi$ we refer here to standard matter fields, which couple to the full four-dimensional metric $g_{\m\n}$. This action can be obtained from that of \emph{non-projectable Ho\v rava Gravity}~\cite{Blas:2009qj} by neglecting operators with more than two derivatives, relevant only at  high energies.
Note that by introducing a preferred foliation, Lorentz symmetry is broken at the local level. The action is invariant under foliation-preserving diffeomorphisms ($T\rightarrow \tilde{T}(T)$, $x^i\rightarrow \tilde{x}^i(x,T)$)
but {\it not} under full four-dimensional diffeomorphisms.

The same action can be recast in covariant form by
promoting the coordinate $T$ to a (timelike) scalar field (the ``khronon'')
and defining a unit-norm, timelike ``\ae ther'' vector field orthogonal to the hypersurfaces of $T=\const$, 
\be \label{khaether}
u_\mu=\frac{\nabla_\mu T}{\sqrt{\nabla^\alpha T\nabla_\alpha T}}\,,
\ee
where we assume a  $+---$ metric signature (as in the following). This allows for writing the action as~\cite{Jacobson:2010mx}
\begin{multline}\label{khaction}
S=-\frac{1}{16\pi G} \int \dd^4x\, \sqrt{-g} \Big[R+\l\;(\nabla_\m u^\m)^2\\+\b\nabla_\m u^\n \nabla_\n u^\m+\a\; a_\m a^\m\Big]+S_{\rm matter}[g_{\mu\nu},\Psi]\,,
\end{multline} 
where $a^\m\equiv u^\n \nabla_\n u^\mu$. Here, LVs are made apparent by the fact that the vector field $\boldsymbol u$ is timelike, i.e., according to the definition \eqref{khaether},
\begin{equation}\label{eq:NormAether}
    u_\mu u^\mu = 1.
\end{equation}

Although still weaker than in matter, LVs in gravity are becoming increasingly constrained, especially by gravitational wave (GW) experiments.
Bounds on the propagation speed of GWs from GW170817 constrain $|\beta|\lesssim 10^{-15}$~\cite{Monitor:2017mdv,Gumrukcuoglu:2017ijh}, which paired with bounds from
solar system experiments also allows for constraining $|\a|\lesssim 10^{-7}$ (with $\lambda$ left unconstrained), \emph{or} $|\a|\lesssim 0.25\times 10^{-4}$ \textit{and} $\lambda\approx \alpha/(1-2\alpha)$~\cite{Will:1993hxu,Will:2014kxa,Blas:2011zd,Bonetti:2015oda,ramos}. 
Measurements of the abundance of primordial elements produced by Big Bang Nucleosynthesis (BBN)
constrain $\l\lesssim 0.1$~\cite{Carroll:2004ai,Yagi:2013ava,Yagi:2013qpa}, with $\l\geq0$ required to ensure absence of ghosts~\cite{Blas:2009qj,Gumrukcuoglu:2017ijh}.
These bounds therefore seem to suggest that $\a$ and $\b$ should be tiny, while $\l$ could still be sizeable. 
Indeed, an additional theoretical constraint -- namely that black holes moving slowly relative to the preferred foliation
remain regular except for their central singularity -- would require  $\a$ and $\b$ to vanish {\it exactly}~\cite{ramos}.

We will refer to the theory with  $\a=\b=0$ and $\l\neq0$ as
{\it minimal Ho\v rava gravity} (mHG) in the following.
Remarkably, all  non-cosmological {\it observables} that have been computed in Ho\v rava gravity reduce to their GR counterparts in the mHG case. %
For instance, the dynamics
in the solar system (i.e. at first post-Newtonian order) 
exactly matches that of GR~\cite{Blas:2011zd,Bonetti:2015oda}. GWs also propagate exactly at the speed of light~\cite{Blas:2011zd}. 
Moreover, static spherically symmetric black holes are  described by the Schwarzschild metric~\cite{Berglund:2012bu}, and so are those moving slowly relative to the preferred foliation~\cite{ramos}. 
The same applies to stars,
for which both static spherically symmetric solutions and 
ones describing slowly moving bodies are characterized by the same (GR) geometry~\cite{Barausse:2019yuk}.
Note that for both stars and black holes the khronon configuration is non-trivial, but does not backreact on the geometry in mHG.
This is quite surprising -- because objects at rest and in motion
are expected to be described by the same metric only in a Lorentz-symmetric theory such as GR, and not (a priori) in a theory with LVs -- and has implications also for
the dynamics of binaries of compact objects and for GW generation.

Indeed, since the geometry of slowly moving stars and black holes is the same as in GR, the ``sensitivities'' -- which parametrize violations of the strong equivalence principle at the leading post-Newtonian (PN) order~\cite{Yagi:2013ava,Yagi:2013qpa,Will:2018ont} -- can be shown to vanish exactly in mHG~\cite{ramos,Barausse:2019yuk}. Therefore, no dipole GW emission
from binaries of compact objects is expected
in mHG~\cite{ramos,Barausse:2019yuk}, unlike for generic $\a,\,\b$ (where this effect was used to test the theory with binary pulsars~\cite{Yagi:2013ava,Yagi:2013qpa}).

A possible caveat 
regarding these experimental bounds 
is that the khronon becomes strongly coupled around the Minkowski and Robertson-Walker geometries in the mHG limit, 
since the scalar field $T$ becomes non-propagating (i.e. its speed diverges) when $\a,\,\b\to0$ and $\l \neq0$~\cite{Blas:2011zd,Kobayashi:2010eh}.
We stress that  strong coupling does not mean that the theory is not viable, but simply that the linearized calculations
on the simple backgrounds
mentioned above may provide incorrect results. However, since
the strong coupling affects the khronon and not the
tensor sector, 
the linear calculation of the speed of GWs (used to compare to the GW170817 observations) is expected to provide trustworthy results.

As for the PN calculations of the solar system dynamics and 
GW generation, it should be noted that
\textit{(i)} the PN 
scheme is an expansion in powers of  $1/c$, and it thus includes non-linear terms; and
\textit{(ii)} the Newtonian/PN dynamics is strongly coupled in GR as well, and yet it gives meaningful results.
Indeed, at leading (Newtonian) order the gravitational field does not propagate in GR (i.e. the equation describing it is elliptic), and propagation only appears at higher PN orders~\cite{Will:1993hxu,Barausse:2013ysa}. We therefore expect
the results from a PN expansion of the field equations to remain valid also in the mHG limit.

Given this wealth of (non-cosmological) observables for which mHG 
provides the same predictions as GR, it is natural to wonder whether mHG and GR may be equivalent, at least in some regimes. Obviously, 
a full equivalence between GR and mHG can be excluded, since the cosmological expansion history is different in the two theories (a fact that is used to constrain $\l$ with BBN data~\cite{Yagi:2013ava,Yagi:2013qpa}), but it may hold
in more specific settings.
 For instance Refs.~\cite{2014arXiv1407.1259L,Bellorin:2010je}, based on a constraint analysis of mHG, claimed that the theory may be equivalent to GR in  vacuum and under asymptotically flat boundary conditions. While suggestive in the light of the ``coincidences''
presented above, this conclusion 
disagrees with that of Ref.~\cite{Henneaux:2009zb}, which solved
the (tertiary) constraint equation of mHG and showed explicitly that the theory cannot be equivalent to GR unless $N=0$ (in which case the metric is degenerate).

In this work, we will therefore attempt to identify (non-cosmological) astrophysical observables for which mHG may differ from GR, focusing on fully non-perturbative calculations, or on ones that involve perturbations over backgrounds  different from the Minkowski and Robertson-Walker geometries (on which the khronon is strongly coupled). In more detail,
in Sec.~\ref{sec:flat} we will review linear perturbations of mHG on flat space.
We will then study the non-linear dynamics of
spherically symmetric collapse (in Sec. \ref{sec:collapse}),
showing that a universal horizon (i.e. a boundary for signals of arbitrary speeds)~\cite{Barausse:2011pu,Blas:2011ni}
naturally forms inside the usual horizon for tensor gravitons and matter.
Nevertheless, the collapse is completely indistinguishable from
GR as far as observable quantities are concerned.
In Sec. \ref{sec:qnm} we will then derive the equations for linear metric perturbations 
over static spherically symmetric black holes, and show that they also coincide with
the GR ones, when focusing on the tensor modes. The scalar mode remains instead strongly coupled (like in flat space), but decouples from the tensor sector.
Our conclusions are finally presented in Sec.~\ref{sec:concl}.

\section{Khronometric theory around flat space} \label{sec:flat}
The dynamics of the action \eqref{khaction} is described in terms of the metric  $g_{\m\n}$ and the \ae ther vector $\mathbf{u}$. The latter is
constrained to be unit-norm and timelike [cf. Eq.~\eqref{khaether}]
and hypersurface orthogonal, i.e. it must have, from the  Fr{\"o}benius theorem, zero vorticity
\begin{equation}\label{eq:frob}
    u_{[\mu} \nabla_\nu u_{\sigma ]}=0\,.
\end{equation}

Since the theory breaks
 boost invariance at the local level, it should propagate additional
 degrees of freedom besides the usual spin-2 graviton field $h_{\m\n}$ of GR.
  Indeed, a generic four-dimensional vector $\boldsymbol{u}$ contains four degrees of freedom -- which can be arranged into a three-dimensional divergence-less vector and two scalars. However, the unit norm \eqref{eq:NormAether} and vorticity \eqref{eq:frob} conditions eliminate three of these degrees of freedom, leaving a single scalar behind (corresponding obviously to the
  khronon scalar field $T$ defining the preferred foliation). 
  
  This can be seen directly at the level of the action by perturbing both the metric and \ae ther around flat space:
\begin{align}
    g_{\m\n}=\eta_{\m\n}+  h_{\m\n},\quad u^\m=(1,\mathbf{0})+v^\m.
\end{align}

Replacing this into the action \eqref{khaction}, we first go to momentum space -- where $\partial_t\equiv i \omega$ and $D_i\equiv i q_i$, with $\omega$ and $q_i$ the frequency and three-momentum respectively -- and perform a $3+1$ decomposition adapted to the foliation orthogonal to the background \ae ther $(1,\mathbf{0})$, i.e. we decompose the metric perturbation and \ae ther as 
\begin{align}
    h_{\m\n}=\begin{pmatrix}
    h_{00}&h_{0i}\\
    h_{0i}&h_{ij}
    \end{pmatrix}, \quad v_\m=(v_0,v_i).
\end{align}
We then split the various quantities in modes
that transform as scalars, vectors and tensors under rotations
\begin{align}
&h_{ij}=\zeta_{ij}+\frac{q_i}{q}X_j +\frac{q_j}{q}X_i+\frac{q_i q_j}{q^2}s_1+\left(\delta_{ij}-\frac{q_i q_j}{q^2}\right)s_2,\\
&v_i=Y_i+\frac{q_i}{q}s_3,\\
&h_{0i}=Z_i+\frac{q_i}{q}s_4,
\end{align}
where $X_i$, $Y_i$ and $Z_i$ are divergenceless vectors -- i.e. $D_i X^i=D_i Y^i=D_i Z^i=0$; $s_i$, $v_0$ and $h_{00}$ are scalars; and $\zeta_{ij}$ is a transverse-traceless tensor -- thus satisfying $D_{i}\zeta^{ij}=D_j\zeta^{ij}=\zeta_i^i=0$.

The two constraints \eqref{eq:NormAether} and \eqref{eq:frob} kill three of these degrees of freedom, as previously mentioned. At the linear level, they impose
\begin{align}
    h_{00}+2v_0=0,\quad \epsilon_{ijk}q^j Y^k=0,
\end{align}
where the second of these conditions is satisfied by setting $Y^k=0$. Once these conditions are enforced, the momentum space Lagrangian for the perturbations, retaining only quadratic terms, becomes
\begin{align}
\nonumber &{\cal L}=- \tfrac{1}{8} (\lambda+ \beta) \omega^2 s_1^2 -  \tfrac{1}{2} (1 + \lambda) \omega^2 s_1^2 -  \lambda q \omega s_1 s_3\\
&+ \tfrac{1}{4} \bigl(- q^2 -  (1 + \beta + 2 \lambda) \omega^2\bigr) s_1^2 -  \tfrac{1}{2} (\lambda+\beta) q \omega s_1 s_3  \nonumber \\ 
& + \tfrac{1}{2} \bigl(- (\lambda+\beta) q^2 -  \alpha \omega^2 \bigr) s_3^2 + q \omega s_1 s_4 -  \alpha \omega^2 s_3 s_4 \\
\nonumber &-  \tfrac{1}{2} \alpha \omega^2 s_4^2 + \tfrac{1}{8} \bigl(q^2 + (1 -\beta) \omega^2\bigr) \zeta_{ab} \zeta^{ab} -  \tfrac{1}{2} q \omega X^{a} Z_{a} \nonumber \\ 
& +  \tfrac{1}{4} (1 -\beta) \omega^2 X_{a} X^{a}  -  \alpha q \omega s_4 v_0 -  \tfrac{1}{2} \alpha q^2 v_0^2 -  \alpha q \omega s_3 v_0\nonumber \\ 
& + \tfrac{1}{4} \bigl((1 - \beta) q^2 - 2 \alpha \omega^2\bigr) Z_{a} Z^{a} + q^2 s_1 v_0,
\end{align}
where  we have omitted a global factor of $G$.

We are left with the task of choosing a suitable gauge. Since the 
action \eqref{khaction} from which we started is covariant (being related to
the ``unitary gauge'' action \eqref{eq:actionHL} by a Stuckelberg transformation),
we need to choose
four gauge conditions. These can  be given as the
requirement that two scalars and one of the three-dimensional divergenceless vectors vanish.
 We choose $s_1=v_0=X^a=0$, and replacing these conditions (as well as those that follow from the equations of motion of these fields) 
 back in the action, we obtain
\begin{align}
{\cal L}&=\frac{1-\beta}{8} \zeta_{ab}\bigl( \omega^2-c_{2}^2 q^2\bigr)  \zeta^{ab} \nonumber \\
&+\frac{(\beta-1)(\lambda+\beta)(2+2\lambda+\beta)}{4(1+\lambda)^2}\tilde{s}(\omega^2 - c_0^2 q^2)\tilde{s}\,,
\end{align}
where we have also rescaled the remaining scalar as $\tilde s=\tfrac{q}{\omega} s_3$.
This is  the Lagrangian of two modes propagating with speeds~\cite{Jacobson:2004ts,Blas:2011zd}
\begin{align}
&c_{2}^2=\frac{1}{1-\beta},\\
&c_0^2=\frac{(\lambda+\beta)(2-\alpha)}{\alpha (1-\beta) (2+3\lambda+\beta)}\label{c0}.
\end{align}

As previously mentioned, we find an extra propagating scalar field with
velocity $c_0$, besides the usual transverse-traceless graviton with velocity $c_2$. Both propagation velocities can be different from the speed of light, although the coincident observation of GW170817 and GRB 170817A
constrains $c_2$ to match $c$ to within about $10^{-15}$ (which in turns bounds $|\beta|\lesssim 10^{-15}$). As for the scalar mode,  cosmic ray observations require $c_0\gtrsim 1$, because
 otherwise ultrahigh energy particles would lose energy to the khronon in a Cherenkov-like cascade~\cite{Elliott:2005va}.
 
Superluminality is of course not
 surprising, since the theory  is not boost-invariant, and thus $c=1$ is not a universal maximum speed. However, although in general the scalar velocity \eqref{c0} is finite, it diverges in the mHG limit,
 if $\lambda\neq0$.
  This is a signal that the linearized expansion breaks down for the dynamics of the scalar field, which is then out of reach of perturbative techniques, while the tensor mode remains healthy. The same conclusion is achieved by performing an identical expansion around FRW space-times or around any maximally symmetric spacetime~\cite{Audren:2013dwa}.
  Note also that this potentially problematic behaviour of the scalar field only appears at low energies. At higher energies the action must be extended by operators with higher number of derivatives, which deform the dispersion relations and lead to a healthy propagating scalar mode.
  
  In light of this ``strong-coupling problem'' for the khronon on flat space, we pursue in the following two distinct calculations in mHG, namely spherically symmetric gravitational collapse and linear perturbations over spherically symmetric static black hole spacetimes. We will aim to assess
  whether the khronon dynamics remains strongly coupled when non-linearities are included in the equations of motion, or when spacetimes more general than Minkowski space are considered

\section{Spherical collapse}\label{sec:collapse} 
Unlike in GR, Birkhoff's theorem does not hold in khronometric theories,
and vacuum spherically symmetric solutions (even when one imposes that they are static and asymptotically
flat) are not unique~\cite{Eling:2006ec,Eling:2006df,Eling:2007xh,Barausse:2011pu,Blas:2011ni}. In more detail, in a given khronometric theory, there exists
a two-parameter family of static, spherically symmetric and asymptotically flat vacuum solutions. One
of the parameters characterizing these solutions is (like in GR) their mass, while
the second parameter regulates the radial tilt of the \ae ther near spatial infinity~\cite{Eling:2006ec,Eling:2006df,Eling:2007xh,Barausse:2011pu,Blas:2011ni}.
In particular, for a given mass, only a specific value of this second parameter yields
solutions that are regular everywhere except for the central $r=0$ curvature singularity. These
are the solutions that are expected to form in gravitational collapse~\cite{Garfinkle:2007bk} and
which are usually referred to as ``black holes'' in the literature~\cite{Eling:2006ec,Barausse:2011pu,Blas:2011ni}.

Although the geometry of 
these black holes is similar to that of the Schwarzschild solution of GR
(with which it actually coincides exactly in the mHG limit),
the existence of the khronon mode has profound implications for their causal structure. As shown in Sec. \ref{sec:flat}, at low energies
Ho\v rava gravity propagates both spin-2 and spin-0 gravitons, whose speeds are generally different from $c$ (i.e. the limiting speed for matter modes).
As a result, different causal boundaries exist for spin-0, spin-2 and matter modes, i.e.  black holes in khronometric theories present spin-0, spin-2 and matter horizons at (generally) distinct locations.

Even more worryingly, when terms of higher order in the (spatial) derivatives are included in the infrared action \eqref{eq:actionHL}, Ho\v rava gravity predicts that the 
 dispersion relations (for both the gravitons and matter) will take the form 
$\omega^2=c_i^2 q^2+A q^4+B q^6$, with $A$ and $B$  constant coefficients and
$c_i$ the species infrared phase velocity. Therefore,
the group velocity $d\omega/dq$ of all species will diverge in the ultraviolet limit, which questions whether it makes sense to talk about event horizons at all. The problem is even more evident in mHG, where the spin-0 propagation speed
diverges already in the infrared limit (cf. Sec.~\ref{sec:flat}).

However, an unavoidable requirement for any physical modes is that they propagate in the future, as defined by the preferred foliation. Therefore, the topology of the hypersurfaces of constant khronon plays a crucial role in defining the spacetime's causal structure. In the infrared black hole solutions of \cite{Barausse:2011pu,Blas:2011ni}, there exists indeed a 
a hypersurface of $T = \const$ that is also a hypersurface of constant radius. Once inside this hypersurface -- which was called the ``universal horizon'' in 
\cite{Barausse:2011pu,Blas:2011ni} -- no modes can escape, even if they propagate at infinite speeds, simply because they need to move in the future direction defined by the background preferred foliation. Note that in the special case of mHG, this universal horizon coincides with the spin-0 horizon, since the khronon propagation speed diverges already in the infrared limit.

Despite their attractive features, as mentioned above, black holes are not the only static and spherically symmetric solutions of khronometric theory. 
Indeed, generic values of the \ae ther tilt parameter
yield solutions that are singular at the spin-0 horizon. In particular,
if the tilt parameter is such that the \ae ther does not present any radial component at spatial infinity, that component vanishes throughout the entire spacetime (i.e. the \ae ther is always parallel to the timelike Killing vector), and the resulting solutions describe the exterior
spacetime of static spherically symmetric stars (whose matter ``covers'' the singularity at the spin-0 horizon)~\cite{Eling:2006df,Eling:2007xh}.

For concreteness, let us examine the special case of 
 mHG, where these static and spherically symmetric vacuum solutions can be obtained analytically and read~\cite{Berglund:2012bu}
\begin{gather}\label{mHGBH1}
	ds^2 = f(r) dt^2 - \frac{B(r)^2}{f(r)} dr^2 - r^2 d\Omega^2, \\  
	u_\alpha dx^\alpha = \frac{1+f(r)A(r)^2}{2A(r)}dt + \frac{B(r)}{2A(r)}\left[\frac{1}{f(r)}-A(r)^2\right]dr\,,\label{mHG_slicing}
\end{gather}
where
\begin{align}\label{ca_soln}
	f(r) &= 1 - \frac{2 G_N M}{r}, \qquad B(r) = 1\,,\\
	A(r) &= \frac{1}{f}\left(-\frac{r_{\aaee}^2}{r^2}+ \sqrt{f+\frac{r_{\aaee}^4}{r^4}}\right)\label{mHGBH2}
\end{align}
The two parameters characterizing each solution are the mass $M$ and the ``radial tilt'' $r_\aaee$. 
These solutions are singular at the universal horizon (which in mHG coincides with the spin-0 horizon, as mentioned above) unless $r_{\aaee} =3^{3/4}G_N M/2$~\cite{Berglund:2012bu}. The latter value  
describes instead a black hole with a regular universal horizon (located at areal radius $r_U=3G_N M/2$), while $r_{\aaee} =0$ describes a static \ae ther 
$\boldsymbol{u}\propto \partial_t$.

Note that Eqs.~\eqref{mHGBH1}--\eqref{mHGBH2}, whatever the value of $r_\aaee$, 
yield $\nabla_\mu u^\mu=0$. If we now express
the class of solutions given by Eqs.~\eqref{mHGBH1}--\eqref{mHGBH2} in
the unitary gauge (where the khronon is used as the time coordinate $T$),
the unit-norm, future directed \ae ther vector $\boldsymbol{u}$ becomes
orthogonal to the preferred foliation $T=$ const, which therefore presents $\K=\nabla_\mu u^\mu=0$. Therefore, Eqs.~\eqref{mHGBH1}--\eqref{mHGBH2}
yield
the Schwarzschild geometry foliated in   maximal (preferred) slices $\K=0$, which
have long been  studied in the context of numerical relativity~\cite{Estabrook:1973ue,Petrich:1985jko,1985ApJ...298...34S,Beig:1997fp,Alcubierre:1138167,Baumgarte:2010ndz}.

In order to ascertain which solution, in the class described by 
Eqs.~\eqref{mHGBH1}--\eqref{mHGBH2}, is produced as the end-point of
gravitational collapse, let us consider the equations of
motion for time dependent configurations, which
can be obtained by varying the action
\eqref{eq:actionHL}.  
Variation with respect to the shift
yields the   momentum constraint $\Ho^i=0$. Variation
with respect to the lapse yields
an equation $\Ho=0$ that reduces to the GR
energy constraint  when $\a,\,\b,\,\l\to0$, but which is {\it not} a priori a constraint equation in khronometric theory. In fact, for generic $\a,\,\b,\,\l$ the resulting equation is
{\it not} a constraint, but corresponds to the khronon's evolution
equation in the covariant formalism of action \eqref{khaction}.
Finally, by varying with respect to $\ga_{ij}$ one obtains 
the evolution equations $\mathcal{E}^{ij}=0$.
We describe matter by a perfect fluid, whose
stress-energy tensor 
\begin{equation}
\SE^{\mu\nu} = (\rho+p)U^\mu U^\nu - p g^{\mu\nu},
\end{equation}
(where $p$, $\rho$ and $\boldsymbol{U}$ are the fluid's pressure, energy density and four-velocity)
is covariantly conserved ($\Dl_\mu \SE^{\mu\nu} = 0$),
since in Eqs.~\eqref{eq:actionHL} and  \eqref{khaction} matter couples only to the four-dimensional metric $g_{\mu\nu}$.
 The explicit form of $\Ho$, $\Ho^i$ and $\mathcal{E}^{ij}$ is given in the Appendix~\ref{app:EOMs_unitgauge} (see also~\cite{Bonetti:2015oda}) for generic khronometric theories, from which the mHG equations
can be obtained by setting $\alpha=\beta=0$.

To simplify the algebra, let us choose spatial coordinates on the preferred slices such that $N^i=0$. Unlike in GR, however, the lapse is 
not a gauge field, i.e. we have already chosen our time coordinate to be the khronon when writing the action~\eqref{eq:actionHL}, and thus
no further conditions can be imposed on $N$. The most generic ansatz that we can write in spherical symmetry is therefore
\begin{gather}\label{ans}
\ga_{ij}\dd x^i \dd x^j = A(T,R)\dd R^2 + R^2 B(T,R) \dd\Omega^2, \\
N = Z(T,R).
\end{gather}
For the matter, we assume that $\rho$ and $p$ are also functions of
$T$ and $R$ alone, and that the three-velocity
is radial, i.e. only $U^R(T,R)$ and $U^T(T,R)$
are non-zero (and related by the normalization condition $U_\alpha U^\alpha=1$).
Note that the class of metrics given by
Eqs.~\eqref{mHGBH1}--\eqref{mHGBH2} can be easily put in the form of
Eq.~\eqref{ans} by performing first a coordinate transformation $t=T+H(r)$ 
(to go to the unitary gauge) and then a further (time dependent) coordinate transformation
$r=R B(T,R)^{1/2}$ to eliminate the shift and render the metric diagonal.
More explicitly, at large radii the solution given by  Eqs.~\eqref{mHGBH1}--\eqref{mHGBH2}
yields
\begin{widetext}
\begin{align}\label{BH1}
Z=&\,1-\frac{G_N M}{R}+\frac{\left(2 k_1-1\right)G_N^2 M^2}{2 R^2}+\frac{\left(-2 k_1^2+2 k_1+2
   k_2-1\right) G_N^3 M^3}{2 R^3}+\nonumber\\&+\frac{-8 G_N M T\, r_{\aaee}^2+4 r_{\aaee}^4+ G_N^4 M^4 \left[8 k_1^3
   -12 k_1^2 +8 k_2 -4 k_1 \left(4 k_2-3\right) -5 \right]}{8  R^4}+{\cal O}\left(\frac{1}{R}\right)^5\,,\\
A=&\,1+\frac{2 G_N M}{R}-\frac{2 \left[\left(k_1+k_2-2\right) G_N^2 M^2\right]}{R^2}+\frac{4 T\, r_{\aaee}^2+G_N^3 M^3 \left(2 k_1^2 -8 k_1 -6 k_2 +8 \right)}{R^3}+{\cal O}\left(\frac{1}{R}\right)^4\,,\\
   B=\,&1+\frac{2 k_1 G_N M}{R}+\frac{\left(k_1^2+2 k_2\right) G_N^2 M^2}{R^2}+\frac{2 k_1 k_2 G_N^3 M^3-2 T\,
   r_{\aaee}^2}{R^3}+{\cal O}\left(\frac{1}{R}\right)^4\,,\label{BHlast}
\end{align}
\end{widetext}
where $k_1$ and $k_2$ are free parameters entering in the coordinate transformation.\footnote{Indeed,
our ansatz \eqref{ans} does not complete fix the gauge,
as it is invariant under a time-independent redefinition of the radius.
This residual gauge freedom arises because
choosing $N_i=0$ does not completely fix the spatial coordinates on $T=$ constant hypersurfaces, but merely ensures that {\it once} a set of spatial coordinates is chosen on some initial  $T=$ constant hypersurface, then the spatial coordinates are fixed in the whole spacetime. The choice of coordinate on the initial slice, however, is arbitrary.} 
Note that even though $Z$, $A$ and $B$ are time dependent, the dependence on time appears at sub-leading order in $1/R$. Moreover, the trace of the extrinsic curvature vanishes, as it did in the original foliation given by Eqs.~\eqref{mHGBH1}--\eqref{mHGBH2}, since
$\K=\nabla_\mu u^\mu$ is a scalar under four-dimensional diffeomorphisms.

 In order to have only first order equations for our system, let us then introduce  $\K_A\equiv {\K_R}^R = -\dl_T A/2A Z $, $\K_B \equiv {\K_\theta}^\theta = {\K_\varphi}^\varphi = -\dl_T B/2B Z$, $D_Z \equiv \dl_R\log Z$, $D_A \equiv \dl_R \log A$ and $D_B \equiv \dl_R \log B$. With these variables,  $\K=-\partial_T\ln\sqrt{\gamma}/N=\K_A+2\K_B$.

With our ansatz, the non-trivial field equations are $\Ho = \Ho^R =\mathcal{E}^{RR}=\mathcal{E}^{\th\th} =0$. As mentioned above, $\Ho =0$ becomes the energy constraint in the GR limit, but is not generically a constraint in khronometric theory.
To check whether $\Ho=0$ is a constraint in mHG, let us 
consider the time derivative of $\Ho$. By using the equations of motions to simplify the expressions, one obtains
\begin{equation}\label{eq:TimeDerivEnConstraint}
\dl_T\Ho = -\frac{\lambda}{R\,AZ} \left[ (2+R\,D)\dl_R \K + r\,\dl^2_R \K \right],
\end{equation}
where $D=2D_Z-D_A/2 + D_B$. Eq.~\eqref{eq:TimeDerivEnConstraint} vanishes either in the GR limit $\lambda = 0$, or when the quantity within brackets is zero. 

Barring the case $\lambda=0$, one therefore has to solve $\dl_T\Ho=0$ (which follows from the original field equation $\Ho=0$) at each time $T$.  Actually, the generic solution to $\dl_T\Ho=0$ is simply
\begin{equation}
\label{eq:MeanCurvatureDerivative}
\dl_R \K =C(T)^2 \frac{A^{1/2}}{R^2 B Z^2},
\end{equation}
where $C(T)$ is an integration constant. 
In a gravitational collapse, \eg~of a star, one requires regularity at the center of the coordinates to obtain 
a physically meaningful solution~\cite{Alcubierre:1138167,Baumgarte:2010ndz}. Necessary conditions for regularity are that
$A$, $Z$, $B$ are finite (and non-vanishing) at $R=0$, and that $\K$ and its radial derivative also remain finite at the center.
 From Eq.~\eqref{eq:MeanCurvatureDerivative}, it is therefore clear that the only way to impose regularity at $r=0$ is to set $C(T)=0$ for any $T$, i.e.
the extrinsic curvature $\K$ must be constant on any given spatial foliation, i.e. $\K(T,R)=k(T)$. Note that a spatially constant trace was also expected from the Hamiltonian analysis of~\cite{Bellorin:2010je,Loll:2014xja}. 

If $k(T)\neq0$, by exploiting time-reparametrization invariance one can set $k(T)=1$.
The evolution equations $\mathcal{E}^{RR}=\mathcal{E}^{\th\th}=0$ and the momentum constraint $\Ho^R=0$ then take the same form as in GR, whereas the equation $\Ho=0$ [which is now a \textit{bona fide} Hamiltonian constraint since $C(T)=0$] contains a term proportional to $\lambda$. However, as noted e.g. by~\cite{Bellorin:2010je}, if asymptotically flat boundary conditions are assumed, at spatial infinity one must necessarily have $\K=k(T)=0$  at all times [cf. also Eqs.~\eqref{BH1}--\eqref{BHlast}]. 
Boundary conditions at spatial infinity that
are not necessarily flat, but which are time-independent, will
yield also $\K=k(T)=0$ at all times (see e.g. \cite{Markovic:1999di} 
for an example of one such GR collapse solution).

Similarly, 
 outgoing boundary conditions at infinity also
imply $\K=k(T)=0$  at all times. This can be seen by noting
that if one imposes $N\approx 1+A_N\exp[i\omega_N(T-R)]/R$ and  
$\sqrt{\gamma}\approx 1+A_\gamma\exp[i\omega_\gamma(T-R)]/R$ (with $A_N$, $A_\gamma$, $\omega_N$, and $\omega_\gamma$ free coefficients), then at large $R$ we find
\begin{equation}
    \dl_R \K \simeq - \frac{A_\ga \om_\ga^2}{R} \exp[\ii \om_\ga (T-R)].
\end{equation}
Requiring that $\partial_R \K=0$ implies $A_\gamma=0$ and thus $\K=0$.

Finally, let us note that if $\K=0$ at all times, the spherical collapse equations and the constraints  become identical to the GR ones, written in the maximal slicing gauge  $\K=0$ [and in our zero-shift ansatz \eqref{ans}]. Since $\K=\nabla_\mu u^\mu$ (with $\boldsymbol{u}$ the unit-norm future-directed 
vector orthogonal to the foliation) is a scalar
under four-dimensional diffeomorphisms, one
can then transform the spherical collapse equations
of mHG into those of GR with maximal time-slicing $\K=0$,
{\it but} more general spatial coordinates (i.e. ones yielding general non-vanishing shift).\footnote{Note that we are changing the spatial coordinates to reinstate the shift,
but {\it not}  the time coordinate, which still coincides with the khronon field (unitary gauge).}

The maximal time-slicing gauge has been extensively used in GR to study gravitational collapse, as it allows for penetrating the black hole horizon~\cite{Estabrook:1973ue,Petrich:1985jko,1985ApJ...298...34S,Beig:1997fp,Alcubierre:1138167,Baumgarte:2010ndz}.
One can therefore 
utilize the results of GR simulations 
(either performed in the maximal slicing gauge, or
transformed to that gauge a posteriori) to gain insight on spherical collapse in mHG.

Indeed, GR collapse simulations in the maximal time-slicing
 found that there exists a ``limiting slice'', i.e. a
limiting hypersurface that the maximal slices
approach at late times~\cite{Estabrook:1973ue,Petrich:1985jko,1985ApJ...298...34S,Beig:1997fp,Alcubierre:1138167,Baumgarte:2010ndz}.
In more detail, the slicing that arises in these simulations outside the collapsing 
sphere turns out to be described 
by the unit-norm future-directed vector $\boldsymbol{u}$ given by Eq.~\eqref{mHG_slicing},
with the parameter $r_\aaee$ asymptotically approaching the critical value $3^{3/4}G_N M/2$. The limiting slice is therefore defined by areal radius $r=3G_N M/2$~\cite{Estabrook:1973ue,Petrich:1985jko,1985ApJ...298...34S,Beig:1997fp,Alcubierre:1138167,Baumgarte:2010ndz}.

While in GR the foliation of the spacetime in time slices has no physical meaning (as it is merely a coordinate effect), the slicing has instead an important physical meaning in mHG, since we are using
the unitary gauge, where the time coordinate coincides with the khronon scalar field. Indeed, the appearance 
of the limiting slice $r=3G_N M/2$ in the GR maximal-slicing collapse simulations
corresponds to the formation of a universal horizon in mHG.\footnote{The correspondence between the appearance of a limiting foliation in GR in the maximal time slicing gauge and the formation of a universal horizon was also noticed in Ref.~\cite{Saravani:2013kva} for the case of Cuscuton theories.
The latter are indeed equivalent to mHG if their scalar potential is quadratic~\cite{Afshordi:2009tt,Bhattacharyya:2016mah}
(even though the equivalence is subtle 
when it comes to the hypersurface orthogonality condition \eqref{eq:frob}, 
as a result of which suitable boundary conditions are required to
obtain an exact equivalence between the two theories~\cite{Afshordi:2009tt}). However,
Ref.~\cite{Saravani:2013kva} worked in the decoupling limit (i.e. 
neglecting the backreaction of the Lorenz-violating field on the geometry)
and with zero potential (in which case the theory is not equivalent to mHG).} This can be understood because in spherical symmetry the universal horizon is, by definition, the outermost
 hypersurface $r= \const$ that is also
orthogonal to $\boldsymbol{u}$ (or equivalently, the outermost hypersurface $r= \const$ that is
also a hypersurface of constant khronon $T= \const$).
We can therefore conclude that spherical collapse in mHG produces ``regular'' black holes, i.e. ones described by Eqs.~\eqref{mHGBH1}--\eqref{mHGBH2}. In particular, no singularity forms at the spin-0/universal horizon.

We  stress, however, that an analysis of the principal part of the fully non-linear spherical collapse equations in generic khronometric theories,
which we present in Appendix \ref{app:char},
shows that the characteristic speed of the scalar mode diverges in the mHG limit.
This suggests that the effect of the khronon on spherical collapse in mHG 
may vanish simply because it satisfies an elliptic equation. It is therefore unclear if gravitational collapse will be the same as in GR when the assumption of spherical symmetry is relaxed. To partially tackle this problem, as well as to assess if moving away from flat space can fix the strong coupling of the khronon
reviewed in Sec.~\ref{sec:flat}, in the next section we will consider linear, but otherwise generic, perturbations of black holes in mHG.

\section{Quasi-normal modes}\label{sec:qnm}

Linear gravitational perturbations of black hole spacetimes in GR have been studied for decades, since the seminal work by Regge, Wheeler and Zerilli for the Schwarzschild geometry~\cite{Regge:1957td,Zerilli:1970se} and  by Teukolsky for the Kerr one~\cite{Teukolsky:1973ha}.
The frequency spectrum of these perturbations, once ingoing/outgoing boundary conditions are imposed at the event horizon/far from the black hole, is
 discrete and consists of complex frequencies. Since the imaginary part of the latter is such that the spectrum modes are exponentially damped (thus
 pointing, in particular, to linear stability of the Schwarzschild and Kerr solutions, at least for non-extremal spins), these modes are usually referred to as quasi-normal modes (QNMs). 
 
 Interestingly, because the Kerr geometry can only depend on two  ``hairs''~\cite{Israel:1967wq,Hawking:1971vc,Carter:1971zc,Robinson:1975bv} (mass and spin\footnote{The electric charge is believed to be zero or extremely small for astrophysical black holes~\cite{Barausse:2014tra}.}), the QNMs frequencies are found to only depend on the same two quantities. This observation has long prompted suggestions to use QNM observations to test the no-hair theorem and thus GR~\cite{Dreyer:2003bv,berti_starinets}, a proposal that the LIGO/Virgo collaboration is starting to tentatively apply to real data~\cite{PhysRevLett.116.221101,LIGOScientific:2019fpa,Isi:2019aib,Abbott:2020jks}, even though really constraining tests will probably have to wait for future detectors~\cite{Berti:2016lat}.

In order to compute QNM frequencies in mHG, let us start from the equations of motion in vacuum derived from the covariant action \eqref{khaction}. From variations of the metric, one obtains
\begin{align}\label{eq:eom_khronometric}
E_{\m\n}=G_{\m\n} -\lambda \SE_{\m\n}^{\rm kh} =0,
\end{align}
where $G_{\m\n}=R_{\m\n}-g_{\m\n}R/2$ and $\SE_{\m\n}^{\rm kh}$ contains the contribution from the khronon:
\begin{align}
\nonumber \SE_{\m\n}^{\rm kh}&= u_\m u_\n u^\sigma\nabla_\sigma\nabla_\rho u^\rho-g_{\m\n} \nabla_\rho u^\rho +\frac{1}{2}g_{\m\n}(\nabla_\sigma u^\sigma)^2\\
&+2\left(g_{(\m}^\rho u_{\n)}-u_{(\m}u_{\n)} u^\rho\right) \nabla_\rho \nabla_\sigma u^\sigma.\label{Tkh}
\end{align}
Variation of the khronon field $T$ yields a scalar 
equation that is equivalent to the covariant conservation of 
 $\SE_{\m\n}^{\rm kh}$, already a consequence of \eqref{eq:eom_khronometric}~\cite{Jacobson:2010mx}.
As such, it does not need to be independently enforced if all of the ten components of  Eq.~\eqref{eq:eom_khronometric} are satisfied. However, 
 we show it here for completeness:
\begin{align}
    \kappa\equiv\lambda \nabla_\m\left[\frac{\left(g^{\m\n}-u^\m u^\n\right)(\nabla_\n \nabla_\sigma u^\sigma)}{\sqrt{\nabla_\a T \nabla^\a T}}\right]=0\,.
\end{align}

Let us now perturb the metric and khronon fields around a curved background geometry, characterized by the pair $\{\overline{g}_{\m\n},\overline{u}_\m\}$: 
\begin{align}\label{lin_ans}
\nonumber &g_{\m\n}=\overline{g}_{\m\n}+\epsilon\  h_{\m\n} +{\cal O}(\epsilon^2),\\
&u_\m=\overline{u}_\m+\epsilon\ v_\m + {\cal O}(\epsilon^2)\,,
\end{align}
where $\epsilon$ is a perturbative parameter
(which  sets the amplitude of the perturbations of the metric and \ae ther/khronon,
which needs to be small for the linear theory to be a good approximation).
In the following, to keep the analysis more general, we will simply assume a spherically symmetric and static background,
given by the  ansatz of Eqs.~\eqref{mHGBH1}--\eqref{mHG_slicing}. To restrict to a black hole background, one can then simply assume the validity of
Eqs.~\eqref{ca_soln}--\eqref{mHGBH2}, with $r_{\aaee} =3^{3/4}G_N M/2$.

Inserting Eq.~\eqref{lin_ans} into Eq. \eqref{eq:eom_khronometric} and expanding  to linear order in $\epsilon$, we obtain the equations of motion for the perturbations in covariant form,
\begin{align}\label{eq:pert_eqs}
\overline{E}_{\m\n}+\epsilon\ \delta E_{\m\n} + {\cal O}(\epsilon^2) = 0,
\end{align}
where $\overline{E}_{\m\n}=0$ is automatic from the choice of background. From now on we will drop the ${\cal O}(\epsilon^2)$ symbol everywhere for notational clarity. Note that the \ae ther field enters Eq.~\eqref{Tkh} both with upper and lower indices. This implies that even if we set $v_\m=0$, we do not trivially recover the same equations for the perturbation as in GR, since there are still non-negligible contributions to $\delta E_{\m\n}$ coming from $u^\m\approx\overline{u}^\m+\epsilon v^\m-\epsilon \overline{u}_\n h^{\m\n}$. Note that this  
signals that the gravitational perturbations ``feel'' the 
presence of a background violating  Lorentz invariance
through the presence of the preferred foliation.

Since the background $\{\overline{g}_{\m\n},\overline{u}_\m\}$ is spherically symmetric, it is convenient to expand the perturbations in spin-weighted spherical harmonics. Using the standard Regge-Wheeler gauge \cite{Regge:1957td} for the metric perturbations and performing a Fourier transform in the time coordinate (exploiting the staticity of the background), we obtain
\begin{align}
h_{\m\n}=\eex^{-i\omega t}\left(h_{\m\n}^{\rm even}+h_{\m\n}^{\rm odd}\sin\theta \, \dl_\th\right)P_\ell(\cos\theta),
\end{align}
where $P_\ell(x)$ is the $\ell$-th Legendre polynomial, with $\ell$ the angular momentum eigenvalue, and
\begin{align}
h_{\m\n}^{\rm even}=\begin{pmatrix}
f(r)H_0^\ell(r) & H_1^\ell(r) & 0 & 0\\
H_1^\ell(r) &\frac{H_2^\ell(r)}{f(r)}& 0 & 0\\
0 & 0 & r^2 K^\ell(r) &0\\
0& 0 &0& r^2 K^\ell(r) \sst
\end{pmatrix},
\end{align}
\begin{align}
h_{\m\n}^{\rm odd}=\begin{pmatrix}
0&0&0&h_0^\ell(r)\\
0&0&0&h_1^\ell(r)\\
0&0&0&0\\
h_0^\ell(r)&h_1^\ell(r)&0&0
\end{pmatrix}.
\end{align}
Here, without loss of generality (thanks to spherical symmetry), we have set the azimuthal number  $m=0$. The functions $H_0(r), H_1(r),H_2(r),K(r),h_0(r)$ and $h_1(r)$, where we have dropped the index $\ell$ to keep the notation compact, characterize the radial profile of the degrees of freedom of the metric perturbations. The perturbation of the \ae ther $v_\m$ depends on that of the khronon field $T$. If we make this explicit in the equations, the expressions quickly become very cumbersome. 
Instead, and equivalently, we choose to write a generic \ae ther vector perturbation
\begin{multline}\label{eq:AnsatzAetherPert}
v_\m= \bigg( \phi_1^\ell(r) ,\phi_2^\ell(r),
2 u_t \phi_3^\ell(r) \dl_\th,0 \bigg)P_\ell(\cos \theta)\eex^{-\ii \om t}, 
\end{multline}
where the factor of $u_t$ is chosen for convenience, since it makes the resulting equations simpler.
Imposing here the unit-norm
and  hypersurface-orthogonality conditions [Eqs.~\eqref{eq:NormAether} and~\eqref{eq:frob}] expanded at linear order in $\epsilon$ allows one to eliminate two of the three free functions appearing in Eq.~\eqref{eq:AnsatzAetherPert}.

Focusing first on the odd part of Eq.~\eqref{eq:pert_eqs}, we find that there are only three potentially independent equations, corresponding to $\de E_{t\theta},\de E_{r\theta}$ and $\de E_{\theta\phi}$. Notice that the perturbation of the \ae ther, arising from the perturbation of a scalar field, has no odd contribution and therefore 
does not appear in the odd sector.

The function $h_0(r)$ can be algebraically solved from the system and, after defining $Q(r)\equiv f(r)h_1(r)/r$, we find that one of the remaining equations implies the other. We are thus left with a single independent equation of the Regge-Wheeler form~\cite{Regge:1957td}
\begin{align}
\frac{\dd^2Q}{\dr_{*}^2}+\left[\omega^2-V_\text{odd}(r)\right]Q=0,
\end{align}
where we have introduced the tortoise coordinate in the usual way, i.e. $\dr/\dr_{*}=f(r)$. The effective potential $V_\text{odd}(r)$ reads
\begin{align}
V_\text{odd}=\frac{\La f(r)}{r^2}+\frac{2 f(r)[f(r)-1]}{r^2}-\frac{f'(r) f(r)}{r},
\end{align}
where $\La=\ell(\ell+1)$.
Replacing the black hole metric [Eq.~\eqref{ca_soln}], this reduces
\emph{exactly} to the potential found in GR for the same Schwarzschild solution. We thus conclude that no differences from GR arise in the equations for odd  perturbations, nor in the QNM frequencies in
this parity sector.

In the even sector, the manipulation of the equations gets more complicated as they now involve the khronon perturbations as well. The full derivation of the equations presented below is shown in Appendix~\ref{appC}, as it is rather lengthy and not particularly enlightening.
In summary, the system is reduced to two second-order equations, for $\phi_3(r)$ and for an additional variable
$\Psi(r)$ defined in   Appendix~\ref{appC}. These two modes 
represent the perturbations of the khronon and  metric, respectively.  The equation for $\Psi$ decouples and reads
\begin{equation}\label{eq:Zerilli}
\frac{\dd^2 \Psi}{\dr_*^2}+\left[\omega^2-V_\text{even}(r)\right]\Psi= 0,
\end{equation}
with the potential
\begin{multline}
V_\text{even}=\frac{f}{r^2 (1+\Lambda -3 f)^2}\bigg[(1+\Lambda)\left(\Lambda(\Lambda-2)+3\right)\\
-3 f\left[(1+\Lambda)^2+3 f (f-1-\Lambda)\right]\bigg].
\end{multline}
This is a wave equation (in Fourier space) and  again it \emph{agrees exactly with the GR result}~\cite{Zerilli:1970se}, once 
specialized to the black hole background [Eq.~\eqref{ca_soln}].
Therefore, the even QNM frequencies for the metric perturbations
coincide with their GR counterparts.

The equation for $\phi_3(r)$, however, remains coupled to $\Psi(r)$, which enters as a source
\begin{align}\label{eqPhi3}
    \phi_3''(r)+W_1(r) \phi_3'(r)+W_0(r)\phi_3(r)=j(r),
\end{align}
with
\begin{align}
j(r)=U_1(r) \Psi'(r)+U_0(r) \Psi(r)
\end{align}
and 
\begin{align}
    &W_1(r)=\frac{-4 A^4 f^2+2 A^2 (5 f+3)-4}{r \left(A^2 f+1\right)^2}+\frac{\omega  \left(2 i-2 i A^2 f\right)}{A^2 f^2+f},\\
   &W_0(r)= \frac{i (3 f+1) \omega  \left(A^2 f-1\right)^3}{f^2 r \left(A^2 f+1\right)^3}-\frac{\omega ^2 \left(A^2 f-1\right)^2}{f^2 \left(A^2 f+1\right)^2}\nonumber\\&\qquad\quad-\frac{4 A^2 \Lambda }{r^2\left(A^2 f +1\right)^2}\,.
\end{align}
The $U_i(r)$ are (very complicated) functions of the geometry, the frequency $\omega$ and the angular momentum $\ell$, and explicit expressions for them are given
 in the Supplemental Material as Mathematica~\cite{Mathematica} files. We have confirmed that this is a general result by looking at the eigensystem of the generalized linear problem, when all equations are taken together: There is no (linear) change of variables which decouples the system into two independent differential equations.

Taking a closer look at our result, it may seem that the khronon field, which was strongly coupled around maximally symmetric spaces (cf. Sec.~\ref{sec:flat}), is now propagating, since Eq.~\eqref{eqPhi3}
has a potential $W_0$ including $\omega^2$, which seems to indicate a finite propagating speed. However, this is just an illusion due to a poor choice of variables, since the equation also contains terms proportional to $\phi_3'(r)$. Performing a change of variables $\phi_3(r)=g(r) \phi(r)$ and choosing $g(r)$ to cancel all terms proportional to $\phi'(r)$, we get
\begin{align}
    &g(r)=C_1 e^{\int_1^r dz\ l(z)},\\
    &l(r)=\frac{i\omega}{f}\frac{A^2f-1}{A^2f+1}+\frac{2+2A^4f^2 -A^2 (3+5f)}{r(A^2 f +1)^2},
\end{align}
where $C_1$ is an integration constant. The equation thus becomes
\begin{align}
    \phi''(r)- V_{\phi}(r)\phi(r)=j(r).
\end{align}
The new potential $V_{\phi}(r)$ has no term proportional to $\omega^2$, whose contribution has been cancelled by that coming from $g(r)$. This corresponds to a  field propagating with infinite speed  (so that $c_\phi^{-2}=0$, with $c_\phi$ the propagation speed). This is analog 
 to the situation in flat space. Therefore, we conclude that the khronon field remains strongly coupled also around black hole geometries (and actually around any static spherically symmetric solution of the class
 described by Eqs.~\eqref{ca_soln}--\eqref{mHGBH2}).
 
 This result can also be confirmed by looking at
the position of the spin-0 horizon for radial khronon modes ($\ell=0$). 
 In more detail, if one
 restores time derivatives in  Eq.~\eqref{eqPhi3} by replacing $\omega \rightarrow i\partial_t$ and takes the eikonal limit -- thus keeping only the highest radial and time derivatives of $\phi_3(r)$ -- we find that Eq.~\eqref{eqPhi3} can be rewritten as
\begin{align}
\mathfrak{g}^{AB}\partial_A\partial_B\phi_3(r)\approx0,    
\end{align}
where the indices $A,B$ run on $\{t,r\}$ and the (inverse) effective metric $\mathfrak{g}$ is given by
\begin{align}
   \mathfrak{g}^{AB} =\begin{pmatrix}
    {1-A^2f}&f({A^2f-1})(1+A^2 f)\\
    f({A^2f-1})(1+A^2 f)&f^2 (1+A^2 f)^2
    \end{pmatrix}.
\end{align}

By computing the null cones of this effective metric, one finds that the apparent horizon for the khronon modes -- which coincides with the event horizon, since the metric is static -- lies at the outermost radius at which 
$\mathfrak{g}^{rr}=0$ (or equivalently $=\mathfrak{g}_{tt}=0$).
Since that location corresponds to $1+A^2(r)f(r)=0$, one can conclude that the spin-0 horizon coincides with the universal horizon, which we recall is the causal boundary for signals of infinite speed.
This again signals that the khronon field  remains strongly coupled even on spherically symmetric and static black hole spacetimes. We stress however that the khronon does not couple directly to matter (and in particular to GW detectors) at tree level.

\section{Conclusions}\label{sec:concl}
In this work, we have explored the possibility of finding astrophysical 
signatures of theories of gravity that violate Lorentz symmetry, focusing on the case of Ho\v rava gravity. 
Because of existing experimental and theoretical constraints, we 
have enforced 
$\alpha=\beta=0$ in the action~\eqref{eq:actionHL} that describes the low energy limit of Ho\v rava gravity, obtaining a theory (for which we coined the
name \emph{minimal Ho\v rava gravity}) depending on only one dimensionless coupling parameter $\lambda$, on which experimental bounds are relatively loose. We have focused 
 on two phenomena that explore both the linear (around curved spacetime) and non-linear dynamics of mHG, i.e., gravitational collapse of 
  spherically symmetric matter configurations and the dynamics of the
  QNMs of the black holes produced by the collapse.

We have found that spherical collapse proceeds exactly as in GR 
as far as the spacetime metric is concerned, but that the khronon field
undergoes a non-trivial dynamical evolution.
In more detail, we have found that the hypersurfaces of constant khronon
follow the same evolution as the maximal time slices $K=0$ of GR, which
are known to asymptote to a limiting slice (corresponding 
to areal radius  $r=3 G_N M/2$) in spherical collapse \cite{Estabrook:1973ue,Petrich:1985jko,1985ApJ...298...34S,Beig:1997fp,Alcubierre:1138167,Baumgarte:2010ndz}. 
 Although in the case of GR the appearance of this limiting slice is just a coordinate effect (since the choice of foliation
 has only practical but not physical meaning), the foliation has a  physical meaning in mHG (where diffeomorphism invariance is broken). We interpret the appearance of the aforementioned limiting slice
 as the formation of a universal horizon (i.e. a causal boundary for signals of arbitray speed). This is a hallmark of black holes in Ho\v rava gravity~\cite{Barausse:2011pu,Blas:2011ni}, and we thus conclude that
 collapse in mHG leads to the formation of a black hole (as opposed to other vacuum solutions of the theory).
 
We have also studied 
QNMs around this family of Lorentz-violating black holes, which are described by the Schwarzschild metric, but which
 present a non-trivial khronon configuration. By using the standard Regge-Wheeler gauge, we have shown that the metric perturbations, both in the even and odd sectors, satisfy the same (linearized) equations as in GR. We have also found that the extra scalar mode of the theory, i.e. the perturbation of the khronon field, remains strongly coupled also around static and spherically symmetric spacetimes.

We therefore conclude that
no (classical) observable deviations from GR arise in either
 spherical collapse or in the spectrum of black hole QNMs, at least if the khronon (which undergoes a non-trivial dynamics) does not couple directly to matter. While direct coupling of the khronon to matter is certainly possible, this would produce large violations of Lorentz invariance in the matter sector, which are tightly constrained by experiments. Another possibility to further test mHG may be provided by cosmology
 (which already places mild constraints on mHG via e.g. BBN)
 and in general by spacetimes which are not asymptotically flat.


\begin{acknowledgments}
We are grateful to  M. Bezares,  L. Lehner, E. Lim and C. Palenzuela  for helpful discussions about spherical collapse, and to D. Blas and S. M. Sibiryakov for numerous conversations about Lorentz-violating gravity.
We acknowledge financial support provided under the European Union's H2020 ERC Consolidator Grant
``GRavity from Astrophysical to Microscopic Scales'' grant agreement no. GRAMS-815673.
\end{acknowledgments}

\appendix
\section{Equations of motion in the unitary gauge}\label{app:EOMs_unitgauge}

The variation of the action~\eqref{eq:actionHL} with respect to the lapse $N$ yields
\begin{equation}\label{eq:energyconstraint}
\begin{split}
\Ho \equiv & {}^{(3)}{R} - (1-\beta) \K_{ij}\K^{ij} + (1+\lambda)\K^2 \\
- & \al a_i a^i -2\al D_i a^i - 8\pi G_N (2-\alpha) N^2 \SE^{00} = 0;
\end{split}
\end{equation}
on the other hand, a variation with respect to the shift $N_i$ gives
\begin{equation}\label{eq:momentumconstraint}
\begin{split}
\Ho^i \equiv & D_j\left(\K^{ij} - \frac{1+\lambda}{1-\beta}\ga^{ij}\K\right) \\
& + 4\pi G_N N \frac{2-\alpha}{1-\beta}(\SE^{0i} + N^i \SE^{00}) = 0;
\end{split}
\end{equation}

finally, variation with respect to the metric $\gamma_{ij}$ gives
\begin{equation}\label{eq:Einsteinequations}
\begin{split}
&\Ein^{ij} \equiv {}^{(3)}{R}^{ij}-\frac{1}{2}{}^{(3)}{R}\ga^{ij} \\
&- \frac{1}{N}D_T\left[(1-\beta)\K^{ij}-(1+\lambda)\ga^{ij}\K\right] \\
&+\frac{2}{N}D_k\left[N^{\left(i\right.}\left((1-\beta)\K^{\left.j\right)k}-(1+\lambda)\K\ga^{\left.j\right)k}\right)\right] \\
&-\frac{1}{2}\ga^{ij}\left[(1-\beta)\K^{kl}\K_{kl}+(1+\lambda)\K^2\right] \\
&+2(1-\beta)\K^{ik}{\K^j}_k-\frac{1}{N}\left(D^i D^jN - \ga^{ij}D_k D^k N\right) \\
&+\al\left(a^ia^j-\frac{1}{2}\ga^{ij}a^2\right)-(1+\beta+2\lambda)\K^{ij}\K \\
& - 4\pi G_N (2-\alpha)( \SE^{ij} - N^i N^j \SE^{00} )= 0,
\end{split}
\end{equation}
where $D_i$ is the covariant derivative  compatible with $\gamma_{ij}$ and $D_T \equiv \dl_T - N_k D^k$.

\section{Characteristic speeds of khronometric theory in spherical symmetry}\label{app:char}
In this Appendix, we  write the 
evolution equations for the metric and khronon [Eqs.~\eqref{eq:Einsteinequations}
and \eqref{eq:energyconstraint}] for the ansatz \eqref{ans} in generic khronometric theories and in spherical symmetry, and compute their characteristic speeds. We also refer the reader to \cite{Garfinkle:2007bk,Bhattacharyya:2015uxt} for more details on 
spherical collapse in  generic khronometric theories. 

By combining with the momentum constraint \eqref{eq:momentumconstraint} and by introducing the variables $X\equiv\partial_T \sqrt{A}/Z$, $Y\equiv \partial_T \sqrt{B}/Z$, $A_R\equiv \partial_R \sqrt{A}$ 
and $B_R\equiv\partial_R \sqrt{B}$, Eqs.~\eqref{eq:Einsteinequations}
and \eqref{eq:energyconstraint} can be put in the first order form
\begin{gather}
\partial_T \boldsymbol{u}+\boldsymbol{M}\cdot \partial_R \boldsymbol{u}= \boldsymbol{S}\label{hypsystem}\\
\partial_R^2 \sqrt{Z}= S_Z \label{eq4A}
\end{gather}
where $\boldsymbol{u}=(X,Y,B_R,A_R)$, $\boldsymbol{M}$ is the characteristic matrix
\begin{widetext}
\begin{equation*}
 \boldsymbol{M}=
\left(
\begin{array}{cccc}
 (\beta +\lambda ) \sqrt{ZA} B  k_1 & 2 (\lambda +1) \sqrt{Z B} A k_1 & \frac{2
   (\alpha -2) (\lambda +1) Z}{\alpha  (\beta -1) (\beta +3 \lambda +2) \sqrt{AB} } & 0 \\
 (\beta +\lambda ) \sqrt{AZ} B  k_2 & 2 (\lambda +1) \sqrt{ZB} A k_2 &
   -\frac{(\alpha -2) (\beta +\lambda ) Z}{\alpha  (\beta -1) (\beta +3 \lambda +2) A} &
   0 \\
 0 & -Z & 0 & 0 \\
 -Z & 0 & 0 & 0 \\
\end{array}
\right)
\end{equation*}
\end{widetext}
while $\boldsymbol{S}$ and $S_Z$ are complicated source terms that depend on
$Z$, $\partial_R Z$, $A$, $B$, $X$, $Y$, $B_R$, $A_R$ and the matter variables.

The characteristic matrix has four eigenvalues 
\begin{gather}
\dot{R}=0\\
\dot{R}= \sqrt{A B Z} \left(k_1 (\beta +\lambda ) \sqrt{B}+2 (\lambda +1) k_2  \sqrt{A}\right)\\
\dot{R}=\pm c_0\frac{Z}{\sqrt{A}}
\end{gather}
where  $k_1$ and $k_2$ are functions of $T$ and $R$, which can be chosen arbitrarily (as they regulate how the momentum constraint is linearly combined with the evolution equations), and $c_0$ is the propagation speed  for the spin-0 modes in Minkowski space [cf. Eq.~\eqref{c0}], 
This means that the sub-system $X,Y,B_R,A_R$ is strongly hyperbolic if $k_1=k_2\ne 0$ and if $c_0$ is real and finite,
while Eq.~\eqref{eq4A} can be solved as an ordinary differential equation
at each time step (provided that suitable boundary conditions are imposed on it).
Note however, as stressed in the main text, that $c_0$ diverges in the mHG limit, signaling a strong-coupling problem.

\section{The linearized field equations for the even-parity sector}\label{appC}

Let us start from the trace-reversed system
\begin{equation}
    \tilde{E}_{\mu\nu} \equiv E_{\mu\nu} - \frac{1}{2} g_{\mu \nu} E_{\alpha\beta} g^{\alpha \beta} = 0\,,
\end{equation}
and use it to compute
the linearized equations $\de \tilde{E}_{\mu\nu}$. In order to simplify them, we make use of the background field equations, and of the unit-norm and hypersurface-orthogonality constraints~\eqref{eq:NormAether}-\eqref{eq:frob}, in order to get rid of $\phi_1$, $\phi_2$ and their derivatives.

In more detail, the seven non-trivial linearized equations have the following structure:
\begin{widetext}
\begin{subequations}
\begin{equation}
    \begin{split}
        \de \tilde{E}_{tt} \propto \, & \lambda\Bigg[C_K^{tt} K +C_{H_0}^{tt} H_0 +C_{H_1}^{tt} H_1 +C_{H_2}^{tt} H_2 + C_{\phi_3}^{tt} \phi _3 
        + C_{K'}^{tt} K'  + C_{H_0'}^{tt} H_0' + C_{H_1'}^{tt} H_1' + C_{H_2'}^{tt} H_2'  + C_{\phi_3'}^{tt} \phi _3' \\
        & + C_{K''}^{tt} K'' + C_{H_0''}^{tt} H_0'' + C_{H_1''}^{tt} H_1'' + C_{H_2''}^{tt} H_2''  + C_{\phi_3''}^{tt} \phi _3'' + C_{\phi_3^{(3)}}^{tt} \phi _3^{(3)}\Bigg]  
        +\frac{(3 f+1) i \omega  H_1}{2 r}-\frac{\omega ^2 H_2}{2} \\
        & + \frac{1}{2} H_0'' f^2+\frac{(f-1) K' f}{2 r}+\frac{(f+3) H_0' f}{4 r} 
        +i \omega  H_1' f-\frac{(f-1) H_2' f}{4 r}-\frac{\Lambda  H_0 f}{2 r^2}-K \omega ^2 = 0,
    \end{split}
\end{equation}
\begin{equation}
    \begin{split}
        \de \tilde{E}_{tr} \propto \, & \lambda\Bigg[C_K^{tr} K +C_{H_0}^{tr} H_0 +C_{H_1}^{tr} H_1 +C_{H_2}^{tr} H_2 + C_{\phi_3}^{tr} \phi _3 
        + C_{K'}^{tr} K'  + C_{H_0'}^{tr} H_0' + C_{H_1'}^{tr} H_1' + C_{H_2'}^{tr} H_2'  + C_{\phi_3'}^{tr} \phi _3' \\
        & + C_{K''}^{tr} K'' + C_{H_0''}^{tr} H_0'' + C_{H_1''}^{tr} H_1'' + C_{H_2''}^{tr} H_2''  + C_{\phi_3''}^{tr} \phi _3'' + C_{\phi_3^{(3)}}^{tr} \phi _3^{(3)}\Bigg]  \\
        & -\frac{i (3 f-1) K \omega }{2 f r}-\frac{H_1 \Lambda }{2 r^2}+\frac{i H_2 \omega }{r}-i \omega  K' = 0,
    \end{split}
\end{equation}
\begin{equation}
    \begin{split}
        \de \tilde{E}_{rr} \propto \, & \lambda\Bigg[C_K^{rr} K +C_{H_0}^{rr} H_0 +C_{H_1}^{rr} H_1 +C_{H_2}^{rr} H_2 + C_{\phi_3}^{rr} \phi _3 
        + C_{K'}^{rr} K'  + C_{H_0'}^{rr} H_0' + C_{H_1'}^{rr} H_1' + C_{H_2'}^{rr} H_2'  + C_{\phi_3'}^{rr} \phi _3' \\
        & + C_{K''}^{rr} K'' + C_{H_0''}^{rr} H_0'' + C_{H_1''}^{rr} H_1'' + C_{H_2''}^{rr} H_2''  + C_{\phi_3''}^{rr} \phi _3'' + C_{\phi_3^{(3)}}^{rr} \phi _3^{(3)}\Bigg]  \\
        & H_2 \left(\frac{\omega ^2}{2 f^2}-\frac{\Lambda }{2 f r^2}\right)+\frac{i (f-1) H_1 \omega }{2 f^2 r}+\frac{3 (f-1) H_0'}{4 f r}-\frac{(3 f+1) H_2'}{4 f r}\\
        & -\frac{i \omega  H_1'}{f}+\frac{(3 f+1) K'}{2 f r}-\frac{H_0''}{2}+K'' = 0,
    \end{split}
\end{equation}
\begin{equation}
    \begin{split}
        \de \tilde{E}_{\th\th} + \frac{\de \tilde{E}_{\phi\phi} }{\sst}\propto \, & \lambda\Bigg[C_K^{\th\th} K +C_{H_0}^{\th\th} H_0 +C_{H_1}^{\th\th} H_1 +C_{H_2}^{\th\th} H_2 + C_{\phi_3}^{\th\th} \phi _3 
        + C_{K'}^{\th\th} K'  + C_{H_0'}^{\th\th} H_0' + C_{H_1'}^{\th\th} H_1'  \\
        & + C_{H_2'}^{\th\th} H_2' + C_{\phi_3'}^{\th\th} \phi _3' + C_{K''}^{\th\th} K'' + C_{H_0''}^{\th\th} H_0'' + C_{H_1''}^{\th\th} H_1'' + C_{H_2''}^{\th\th} H_2''  + C_{\phi_3''}^{\th\th} \phi _3'' + C_{\phi_3^{(3)}}^{\th\th} \phi _3^{(3)}\Bigg]  \\
        & -f r H_0'-f r H_2'+f r^2 K''+(3 f r+r) K'+K \left(\frac{r^2 \omega ^2}{f}-\Lambda +2\right) \\
        & +\frac{H_0 \Lambda }{2}+H_2 \left(-\frac{\Lambda }{2}-2\right)-2 i H_1 r \omega = 0,
    \end{split}
\end{equation}
\begin{equation}
\begin{split}
    \de \tilde{E}_{t \theta} \propto \, & \lambda  \Bigg[H_0 \left(\frac{(3 f+1) \left(A^2 f+1\right) \left(A^2 f-1\right)^3}{32
   A^4 f^2 r}+\frac{i \omega  \left(A^4 f^2-1\right)^2}{32 A^4 f^2}\right) \\
   & +H_1
   \left(\frac{\left(A^2 f+1\right)^2 \left(3 A^4 f^3+\left(A^2-12\right) A^2
   f^2+\left(3-4 A^2\right) f+1\right)}{16 A^4 f^2 r}+\frac{i \omega  \left(A^2
   f+1\right) \left(A^2 f-1\right)^3}{16 A^4 f^2}\right) \\
   & +H_2 \left(\frac{(3 f+1)
   \left(A^2 f+1\right) \left(A^2 f-1\right)^3}{32 A^4 f^2 r}+\frac{i \omega  \left(A^2
   f+1\right)^2 \left(A^4 f^2-6 A^2 f+1\right)}{32 A^4 f^2}\right) \\
   & +\frac{\left(-A^8 f^4-6 A^6 f^3+6 A^2 f+1\right) H_0'}{32 A^4 f}+\frac{\left(A^4
   f^2-1\right) K'}{4 A^2} \\
   & +\phi _3 \left(\frac{i (3 f+1) \omega  \left(A^2 f+1\right)
   \left(A^2 f-1\right)^3}{8 A^4 f^2 r}-\frac{\omega ^2 \left(A^4 f^2-1\right)^2}{8 A^4
   f^2}-\frac{\Lambda  \left(A^2 f+1\right)^2}{2 A^2 r^2}\right) \\
   & +\phi _3'
   \left(-\frac{\left(2 A^4 f^2-A^2 (5 f+3)+2\right) \left(A^2 f+1\right)^2}{4 A^4
   r}-\frac{i \omega  \left(A^2 f-1\right) \left(A^2 f+1\right)^3}{4 A^4
   f}\right) \\
   & -\frac{\left(A^2 f+1\right)^4 H_1'}{16 A^4 f}-\frac{\left(A^2 f-1\right)
   \left(A^2 f+1\right)^3 H_2'}{32 A^4 f}-\frac{i K \omega  \left(A^2 f+1\right)^2}{4 A^2
   f}+\frac{\left(A^2 f+1\right)^4 \phi _3''}{8 A^4}\Bigg] \\
   & +\frac{(f-1) H_1}{2 r}-\frac{f
   H_1'}{2}-\frac{1}{2} i H_2 \omega -\frac{1}{2} i K \omega = 0,
\end{split}
\end{equation}
\begin{equation}
    \begin{split}
        \de \tilde{E}_{r\theta} \propto \, & \lambda  \Bigg[H_0 \left(-\frac{(3 f+1) \left(A^2 f-1\right)^4}{32 A^4 f^3 r}-\frac{i \omega  \left(A^2 f+1\right) \left(A^2 f-1\right)^3}{32 A^4 f^3}\right) \\
        & +H_1 \left(\frac{-3 A^8 f^5-\left(A^2-12\right) A^6 f^4+4 A^6 f^3-12 A^2 f^2+\left(3-4 A^2\right) f+1}{16 A^4 f^3 r}-\frac{i \omega  \left(A^2 f-1\right)^4}{16 A^4 f^3}\right) \\
        & +H_2 \left(-\frac{(3 f+1) \left(A^2 f-1\right)^4}{32 A^4 f^3 r}-\frac{i \omega  \left(A^8 f^4-6 A^6 f^3+6 A^2 f-1\right)}{32 A^4 f^3}\right) +\frac{\left(A^4 f^2-1\right)^2 H_2'}{32 A^4 f^2}\\
        & +\frac{\left(A^2 f-1\right) \left(A^2 f+1\right)^3 H_1'}{16 A^4 f^2}+\frac{\left(A^2 f-1\right)^2 \left(A^4 f^2+6 A^2 f+1\right) H_0'}{32 A^4 f^2}+\frac{i K \omega  \left(A^4 f^2-1\right)}{4 A^2 f^2} \\
        & +\phi _3 \left(\frac{\Lambda  \left(A^4 f^2-1\right)}{2 A^2 f r^2}-\frac{i (3 f+1) \omega  \left(A^2 f-1\right)^4}{8 A^4 f^3 r}+\frac{\omega ^2 \left(A^2 f+1\right) \left(A^2 f-1\right)^3}{8 A^4 f^3}\right) \\
        & +\phi _3' \left(\frac{2 A^8 f^4-A^6 f^2 (5 f+3)+A^2 (5 f+3)-2}{4 A^4 f r}+\frac{i \omega  \left(A^4 f^2-1\right)^2}{4 A^4 f^2}\right)-\frac{\left(A^2 f-1\right)^2 K'}{4 A^2 f} \\
        & -\frac{\left(A^2 f-1\right) \left(A^2 f+1\right)^3 \phi _3''}{8 A^4 f}\Bigg]+\frac{(3 f-1) H_0}{4 f r}-\frac{(f+1) H_2}{4 f r}-\frac{i H_1 \omega }{2 f}-\frac{H_0'}{2}+\frac{K'}{2} = 0,
    \end{split}
\end{equation}
\begin{equation}
    \de \tilde{E}_{\theta\phi} \propto H_0 - H_2 = 0\,.
\end{equation}
\end{subequations}
\end{widetext}
The explicit expressions for the coefficients $C^{ij}_k$ are given in the Supplemental Material as Mathematica~\cite{Mathematica} files.

Let us notice that these seven equations contain only five independent variables $H_0,H_1,H_2,K$ and $\phi_3$. This seems to imply that the system may be over-determined. This turns out not to be the case, since some of these equations are redundant due to the Bianchi identity.

In more detail, from diffeomorphism invariance of the covariant gravitational action \eqref{khaction} (without the matter contribution) one obtains the generalized Bianchi identity~\cite{ramos}
\begin{align}
    \nabla_\m E^{\m\n}=-\frac{\kappa}{2}\sqrt{\nabla_\a T \nabla^\a T} \  u^\n\,.
\end{align}
Taking linear combinations to cancel out the explicit dependence on $T$ and $\kappa$ and performing trivial manipulations, one can then write the identity
\begin{align}
  \nabla_\n\left(E^{\m\n}u^\a - E^{\a\n}u^\m\right)= E^{\m\n}\nabla_\n u^\a - E^{\a\n}\nabla_\n u^\m\,,
\end{align}
which can be used to show that two of the seven equations can be eliminated from the system without loss of generality. This can also be seen by direct manipulation of the equations of motion, as we will now show.

From $\de \tilde{E}_{\theta\phi} = 0$, we obtain 
\begin{equation}
    H_2(r) = H_0(r),
\end{equation}
which allows us to get rid of $H_2$ completely. 
The structure of the remaining equations is then the following:
on the one hand,
the equations $\de \tilde{E}_{tt},\, \de \tilde{E}_{tr},\,\de \tilde{E}_{rr},\,\de \tilde{E}_{\theta\theta}$ contain up to second derivatives of the metric perturbations and up to third derivatives of $\phi_3(r)$\footnote{Third radial derivatives appear after imposing the hypersurface-orthogonality condition, Eq.~\eqref{eq:frob}.}; on the other hand, in $\de \tilde{E}_{t\theta}$ and $\de \tilde{E}_{r\theta}$ one can find up to first derivatives of the metric perturbations and up to second derivatives of $\phi_3(r)$.
Thus, from $\de \tilde{E}_{t\theta}=0$ and $\de \tilde{E}_{r\theta}=0$ we can solve algebraically for $H_1'(r)$ and $\phi_3''(r)$. This defines an equation for the scalar field, which  we denote by $F_\phi=0$. The next step is to use the expressions for $H_1'(r)$ and $\phi_3''(r)$ (and their derivatives) to eliminate $H_1'$, $H_1''$, $\phi_3''$ and $\phi_3^{(3)}$ from the rest of the equations. By doing so, we obtain $\tilde{E}_{rr} \propto \tilde{E}_{tt}$.

We then solve $\de E_{tt}=0$, $\de E_{rt}=0$ and $\de E_{\th\th}=0$ and get  algebraic expressions for $H_0''(r)$, $K''(r)$ and $H_1(r)$, which take the schematic form
\begin{align}
    &F_0\equiv H_0'' - d_1 H_0' + d_2 K' + d_3 H_0 + d_4 K=0, \label{eq:H0second}\\
    &F_K\equiv K''   - d_5 H_0' + d_6 K' + d_7 H_0 + d_8 K = 0, \label{eq:Ksecond}\\
    &H_1   - d_9 H_0' + d_{10} K' + d_{11} H_0 + d_{12} K=0, \label{eq:H1zero}
\end{align}
where the $d_i$ are functions of $r$, $\om$ and $\La$. We have checked that the derivative of Eq.~\eqref{eq:H1zero} coincides with the previous analytic solution that we had found for $H_1'$, so Eq.~\eqref{eq:H1zero} is redundant. We are thus left with three independent equations, corresponding to $F_\phi=0$, $F_0=0$ and $F_K=0$, which depend only on three variables $\phi_3$, $H_0$ and $K$, with the scalar field present only in $F_\phi$. Moreover, both $F_0$ and $F_k$ are independent of $\lambda$, and $F_\phi$ contains only first derivatives of $H_0$ and $K$.

In the GR limit, $\lambda\to 0$, the dependence on $\phi_3$  also disappears from $F_\phi$. In that case, compatibility of the system would require that one of the equations is redundant. Note that this must be the case since we know that in the GR limit
the energy constraint is re-instated, cf. Eq.~\eqref{eq:TimeDerivEnConstraint}.
Since $F_\phi$ reduces to a first order equation in the GR limit, it can be used, upon substitution into the other equations, to reduce the whole system to two first-order equations relating $H_0$ and $K$:
\begin{widetext}
\begin{subequations}
\label{eq:H0Kprime}
\begin{equation}
\begin{split}
    &-\frac{\left(f^2 (-\Lambda )+f \left(\Lambda ^2-2 \Lambda +14 r^2 \omega ^2\right)+\Lambda -2 \Lambda  r^2 \omega ^2-6 r^2 \omega ^2\right)}{f r \left(f \Lambda -\Lambda +4 r^2 \omega ^2\right)}H_0 \\
    &+\frac{\left(f^2 \left(\Lambda ^2-2 \Lambda +9 r^2 \omega ^2\right)-2 f (2 \Lambda +1) r^2 \omega ^2+4 r^4 \omega ^4+r^2 \omega ^2\right)}{f^2 r \left(f \Lambda -\Lambda +4 r^2 \omega ^2\right)}K+H_0' = 0,
\end{split}
\end{equation}
\begin{equation}
    \frac{H_0 \left(-2 f \Lambda +\Lambda ^2+4 r^2 \omega ^2\right)}{-f \Lambda  r-4 r^3 \omega ^2+\Lambda  r}+\frac{K \left(f \left(\Lambda ^2-2 \Lambda +6 r^2 \omega ^2\right)-2 (\Lambda +1) r^2 \omega ^2\right)}{f r \left(f \Lambda -\Lambda +4 r^2 \omega ^2\right)}+K' = 0.
\end{equation}
\end{subequations}
\end{widetext}
Since Eqs.~\eqref{eq:H0second}-\eqref{eq:Ksecond} can be shown to be independent of $\lambda$, Eq.~\eqref{eq:H0Kprime}
 also holds in the general case, as can be checked by direct substitution.

A last simplification occurs by introducing the same variable transformation as in \cite{Zerilli:1970se,Sago:2002fe}, given by
\begin{align}
    & K = \frac{\Lambda (1+\Lambda) - 3(2+\Lambda)f + 6f^2}{2 r \left(1+\Lambda -3 f\right)}\Psi+ f \Psi', \\
    & \begin{multlined}
        H_0 = -\frac{1+\Lambda - 3 \Lambda f +3 f^2}{2(1+\Lambda-3f)}\Psi'  \\
        +\left[\frac{1+\Lambda -3 f}{6r}+\frac{(\Lambda-2)^2 (1+\Lambda)}{3r(1+\Lambda-3f)^2}-\frac{r \omega^2}{f}\right]\Psi.
        \end{multlined}
\end{align}
After performing this transformation, Eq.~\eqref{eq:H0Kprime} reduces to 
the simple equation
\begin{equation}
\frac{\dd^2 \Psi}{\dr_*^2}+\left[\omega^2-V_\text{even}(r)\right]\Psi= 0,
\end{equation}
with potential
\begin{multline}
V_\text{even}=\frac{f}{r^2 (1+\Lambda -3 f)^2}\bigg[(1+\Lambda)\left(\Lambda(\Lambda-2)+3\right)\\
-3 f\left[(1+\Lambda)^2+3 f (f-1-\Lambda)\right]\bigg].
\end{multline}

However, in the general case of non-vanishing $\lambda$, the third equation $F_\phi=0$ remains independent and serves as the equation of motion for the scalar field:
\begin{align}
    \phi_3''(r)+W_1(r) \phi_3'(r)+W_0(r)\phi_3(r)=j(r),
\end{align}
with
\begin{align}
j(r)=U_1(r) \Psi'(r)+U_0(r) \Psi(r).
\end{align}
The explicit forms of the functions $W_i(r)$ are
\begin{widetext}
\begin{align}
    &W_1(r)=\frac{-4 A^4 f^2+2 A^2 (5 f+3)-4}{r \left(A^2 f+1\right)^2}+\frac{\omega  \left(2 i-2 i A^2 f\right)}{A^2 f^2+f},\\
   &W_0(r)= \frac{i (3 f+1) \omega  \left(A^2 f-1\right)^3}{f^2 r \left(A^2 f+1\right)^3}-\frac{\omega ^2 \left(A^2 f-1\right)^2}{f^2 \left(A^2 f+1\right)^2}-\frac{4 A^2 \Lambda }{r^2\left(A^2 f +1\right)^2},
\end{align}
while $U_0(r)$ and $U_1(r)$ are included in the Supplemental Material as Mathematica~\cite{Mathematica} files.
\end{widetext}

\bibliographystyle{apsrev4-1}

\bibliography{shortbib}

\begin{thebibliography}{74}%
\makeatletter
\providecommand \@ifxundefined [1]{%
 \@ifx{#1\undefined}
}%
\providecommand \@ifnum [1]{%
 \ifnum #1\expandafter \@firstoftwo
 \else \expandafter \@secondoftwo
 \fi
}%
\providecommand \@ifx [1]{%
 \ifx #1\expandafter \@firstoftwo
 \else \expandafter \@secondoftwo
 \fi
}%
\providecommand \natexlab [1]{#1}%
\providecommand \enquote  [1]{``#1''}%
\providecommand \bibnamefont  [1]{#1}%
\providecommand \bibfnamefont [1]{#1}%
\providecommand \citenamefont [1]{#1}%
\providecommand \href@noop [0]{\@secondoftwo}%
\providecommand \href [0]{\begingroup \@sanitize@url \@href}%
\providecommand \@href[1]{\@@startlink{#1}\@@href}%
\providecommand \@@href[1]{\endgroup#1\@@endlink}%
\providecommand \@sanitize@url [0]{\catcode `\\12\catcode `\$12\catcode
  `\&12\catcode `\#12\catcode `\^12\catcode `\_12\catcode `\%12\relax}%
\providecommand \@@startlink[1]{}%
\providecommand \@@endlink[0]{}%
\providecommand \url  [0]{\begingroup\@sanitize@url \@url }%
\providecommand \@url [1]{\endgroup\@href {#1}{\urlprefix }}%
\providecommand \urlprefix  [0]{URL }%
\providecommand \Eprint [0]{\href }%
\providecommand \doibase [0]{http://dx.doi.org/}%
\providecommand \selectlanguage [0]{\@gobble}%
\providecommand \bibinfo  [0]{\@secondoftwo}%
\providecommand \bibfield  [0]{\@secondoftwo}%
\providecommand \translation [1]{[#1]}%
\providecommand \BibitemOpen [0]{}%
\providecommand \bibitemStop [0]{}%
\providecommand \bibitemNoStop [0]{.\EOS\space}%
\providecommand \EOS [0]{\spacefactor3000\relax}%
\providecommand \BibitemShut  [1]{\csname bibitem#1\endcsname}%
\let\auto@bib@innerbib\@empty
\bibitem [{\citenamefont {Kostelecky}(2004)}]{Kostelecky:2003fs}%
  \BibitemOpen
  \bibfield  {author} {\bibinfo {author} {\bibfnamefont {V.~A.}\ \bibnamefont
  {Kostelecky}},\ }\href {\doibase 10.1103/PhysRevD.69.105009} {\bibfield
  {journal} {\bibinfo  {journal} {Phys. Rev.}\ }\textbf {\bibinfo {volume}
  {D69}},\ \bibinfo {pages} {105009} (\bibinfo {year} {2004})},\ \Eprint
  {http://arxiv.org/abs/hep-th/0312310} {arXiv:hep-th/0312310 [hep-th]}
  \BibitemShut {NoStop}%
\bibitem [{\citenamefont {Kostelecky}\ and\ \citenamefont
  {Russell}(2011)}]{Kostelecky:2008ts}%
  \BibitemOpen
  \bibfield  {author} {\bibinfo {author} {\bibfnamefont {V.~A.}\ \bibnamefont
  {Kostelecky}}\ and\ \bibinfo {author} {\bibfnamefont {N.}~\bibnamefont
  {Russell}},\ }\href {\doibase 10.1103/RevModPhys.83.11} {\bibfield  {journal}
  {\bibinfo  {journal} {Rev. Mod. Phys.}\ }\textbf {\bibinfo {volume} {83}},\
  \bibinfo {pages} {11} (\bibinfo {year} {2011})},\ \Eprint
  {http://arxiv.org/abs/0801.0287} {arXiv:0801.0287 [hep-ph]} \BibitemShut
  {NoStop}%
\bibitem [{\citenamefont {Mattingly}(2005)}]{Mattingly:2005re}%
  \BibitemOpen
  \bibfield  {author} {\bibinfo {author} {\bibfnamefont {D.}~\bibnamefont
  {Mattingly}},\ }\href {\doibase 10.12942/lrr-2005-5} {\bibfield  {journal}
  {\bibinfo  {journal} {Living Rev. Rel.}\ }\textbf {\bibinfo {volume} {8}},\
  \bibinfo {pages} {5} (\bibinfo {year} {2005})},\ \Eprint
  {http://arxiv.org/abs/gr-qc/0502097} {arXiv:gr-qc/0502097 [gr-qc]}
  \BibitemShut {NoStop}%
\bibitem [{\citenamefont {Jacobson}\ \emph {et~al.}(2006)\citenamefont
  {Jacobson}, \citenamefont {Liberati},\ and\ \citenamefont
  {Mattingly}}]{Jacobson:2005bg}%
  \BibitemOpen
  \bibfield  {author} {\bibinfo {author} {\bibfnamefont {T.}~\bibnamefont
  {Jacobson}}, \bibinfo {author} {\bibfnamefont {S.}~\bibnamefont {Liberati}},
  \ and\ \bibinfo {author} {\bibfnamefont {D.}~\bibnamefont {Mattingly}},\
  }\href {\doibase 10.1016/j.aop.2005.06.004} {\bibfield  {journal} {\bibinfo
  {journal} {Annals Phys.}\ }\textbf {\bibinfo {volume} {321}},\ \bibinfo
  {pages} {150} (\bibinfo {year} {2006})},\ \Eprint
  {http://arxiv.org/abs/astro-ph/0505267} {arXiv:astro-ph/0505267 [astro-ph]}
  \BibitemShut {NoStop}%
\bibitem [{\citenamefont {Jacobson}(2007)}]{Jacobson:2008aj}%
  \BibitemOpen
  \bibfield  {author} {\bibinfo {author} {\bibfnamefont {T.}~\bibnamefont
  {Jacobson}},\ }\bibfield  {booktitle} {\emph {\bibinfo {booktitle}
  {{Proceedings, Workshop on From quantum to emergent gravity: Theory and
  phenomenology (QG-Ph): Trieste, Italy, June 11-15, 2007}}},\ }\href@noop {}
  {\bibfield  {journal} {\bibinfo  {journal} {PoS}\ }\textbf {\bibinfo {volume}
  {QG-PH}},\ \bibinfo {pages} {020} (\bibinfo {year} {2007})},\ \Eprint
  {http://arxiv.org/abs/0801.1547} {arXiv:0801.1547 [gr-qc]} \BibitemShut
  {NoStop}%
\bibitem [{\citenamefont {Liberati}(2013)}]{Liberati:2013xla}%
  \BibitemOpen
  \bibfield  {author} {\bibinfo {author} {\bibfnamefont {S.}~\bibnamefont
  {Liberati}},\ }\href {\doibase 10.1088/0264-9381/30/13/133001} {\bibfield
  {journal} {\bibinfo  {journal} {Class.Quant.Grav.}\ }\textbf {\bibinfo
  {volume} {30}},\ \bibinfo {pages} {133001} (\bibinfo {year} {2013})},\
  \Eprint {http://arxiv.org/abs/1304.5795} {arXiv:1304.5795 [gr-qc]}
  \BibitemShut {NoStop}%
\bibitem [{\citenamefont {Kostelecky}\ and\ \citenamefont
  {Tasson}(2011)}]{Kostelecky:2010ze}%
  \BibitemOpen
  \bibfield  {author} {\bibinfo {author} {\bibfnamefont {A.~V.}\ \bibnamefont
  {Kostelecky}}\ and\ \bibinfo {author} {\bibfnamefont {J.~D.}\ \bibnamefont
  {Tasson}},\ }\href {\doibase 10.1103/PhysRevD.83.016013} {\bibfield
  {journal} {\bibinfo  {journal} {Phys. Rev.}\ }\textbf {\bibinfo {volume}
  {D83}},\ \bibinfo {pages} {016013} (\bibinfo {year} {2011})},\ \Eprint
  {http://arxiv.org/abs/1006.4106} {arXiv:1006.4106 [gr-qc]} \BibitemShut
  {NoStop}%
\bibitem [{\citenamefont {Ho\v{r}ava}(2009)}]{Horava:2009uw}%
  \BibitemOpen
  \bibfield  {author} {\bibinfo {author} {\bibfnamefont {P.}~\bibnamefont
  {Ho\v{r}ava}},\ }\href {\doibase 10.1103/PhysRevD.79.084008} {\bibfield
  {journal} {\bibinfo  {journal} {Phys. Rev.}\ }\textbf {\bibinfo {volume}
  {D79}},\ \bibinfo {pages} {084008} (\bibinfo {year} {2009})},\ \Eprint
  {http://arxiv.org/abs/0901.3775} {arXiv:0901.3775 [hep-th]} \BibitemShut
  {NoStop}%
\bibitem [{\citenamefont {Barvinsky}\ \emph {et~al.}(2016)\citenamefont
  {Barvinsky}, \citenamefont {Blas}, \citenamefont {Herrero-Valea},
  \citenamefont {Sibiryakov},\ and\ \citenamefont
  {Steinwachs}}]{Barvinsky:2015kil}%
  \BibitemOpen
  \bibfield  {author} {\bibinfo {author} {\bibfnamefont {A.~O.}\ \bibnamefont
  {Barvinsky}}, \bibinfo {author} {\bibfnamefont {D.}~\bibnamefont {Blas}},
  \bibinfo {author} {\bibfnamefont {M.}~\bibnamefont {Herrero-Valea}}, \bibinfo
  {author} {\bibfnamefont {S.~M.}\ \bibnamefont {Sibiryakov}}, \ and\ \bibinfo
  {author} {\bibfnamefont {C.~F.}\ \bibnamefont {Steinwachs}},\ }\href
  {\doibase 10.1103/PhysRevD.93.064022} {\bibfield  {journal} {\bibinfo
  {journal} {Phys. Rev.}\ }\textbf {\bibinfo {volume} {D93}},\ \bibinfo {pages}
  {064022} (\bibinfo {year} {2016})},\ \Eprint
  {http://arxiv.org/abs/1512.02250} {arXiv:1512.02250 [hep-th]} \BibitemShut
  {NoStop}%
\bibitem [{\citenamefont {Chadha}\ and\ \citenamefont
  {Nielsen}(1983)}]{Chadha:1982qq}%
  \BibitemOpen
  \bibfield  {author} {\bibinfo {author} {\bibfnamefont {S.}~\bibnamefont
  {Chadha}}\ and\ \bibinfo {author} {\bibfnamefont {H.~B.}\ \bibnamefont
  {Nielsen}},\ }\href {\doibase 10.1016/0550-3213(83)90081-0} {\bibfield
  {journal} {\bibinfo  {journal} {Nucl. Phys.}\ }\textbf {\bibinfo {volume}
  {B217}},\ \bibinfo {pages} {125} (\bibinfo {year} {1983})}\BibitemShut
  {NoStop}%
\bibitem [{\citenamefont {Bednik}\ \emph {et~al.}(2013)\citenamefont {Bednik},
  \citenamefont {Pujolas},\ and\ \citenamefont {Sibiryakov}}]{Bednik:2013nxa}%
  \BibitemOpen
  \bibfield  {author} {\bibinfo {author} {\bibfnamefont {G.}~\bibnamefont
  {Bednik}}, \bibinfo {author} {\bibfnamefont {O.}~\bibnamefont {Pujolas}}, \
  and\ \bibinfo {author} {\bibfnamefont {S.}~\bibnamefont {Sibiryakov}},\
  }\href {\doibase 10.1007/JHEP11(2013)064} {\bibfield  {journal} {\bibinfo
  {journal} {JHEP}\ }\textbf {\bibinfo {volume} {11}},\ \bibinfo {pages} {064}
  (\bibinfo {year} {2013})},\ \Eprint {http://arxiv.org/abs/1305.0011}
  {arXiv:1305.0011 [hep-th]} \BibitemShut {NoStop}%
\bibitem [{\citenamefont {Barvinsky}\ \emph {et~al.}(2017)\citenamefont
  {Barvinsky}, \citenamefont {Blas}, \citenamefont {Herrero-Valea},
  \citenamefont {Sibiryakov},\ and\ \citenamefont
  {Steinwachs}}]{Barvinsky:2017kob}%
  \BibitemOpen
  \bibfield  {author} {\bibinfo {author} {\bibfnamefont {A.~O.}\ \bibnamefont
  {Barvinsky}}, \bibinfo {author} {\bibfnamefont {D.}~\bibnamefont {Blas}},
  \bibinfo {author} {\bibfnamefont {M.}~\bibnamefont {Herrero-Valea}}, \bibinfo
  {author} {\bibfnamefont {S.~M.}\ \bibnamefont {Sibiryakov}}, \ and\ \bibinfo
  {author} {\bibfnamefont {C.~F.}\ \bibnamefont {Steinwachs}},\ }\href
  {\doibase 10.1103/PhysRevLett.119.211301} {\bibfield  {journal} {\bibinfo
  {journal} {Phys. Rev. Lett.}\ }\textbf {\bibinfo {volume} {119}},\ \bibinfo
  {pages} {211301} (\bibinfo {year} {2017})},\ \Eprint
  {http://arxiv.org/abs/1706.06809} {arXiv:1706.06809 [hep-th]} \BibitemShut
  {NoStop}%
\bibitem [{\citenamefont {Barvinsky}\ \emph {et~al.}(2019)\citenamefont
  {Barvinsky}, \citenamefont {Herrero-Valea},\ and\ \citenamefont
  {Sibiryakov}}]{Barvinsky:2019rwn}%
  \BibitemOpen
  \bibfield  {author} {\bibinfo {author} {\bibfnamefont {A.~O.}\ \bibnamefont
  {Barvinsky}}, \bibinfo {author} {\bibfnamefont {M.}~\bibnamefont
  {Herrero-Valea}}, \ and\ \bibinfo {author} {\bibfnamefont {S.~M.}\
  \bibnamefont {Sibiryakov}},\ }\href {\doibase 10.1103/PhysRevD.100.026012}
  {\bibfield  {journal} {\bibinfo  {journal} {Phys. Rev. D}\ }\textbf {\bibinfo
  {volume} {100}},\ \bibinfo {pages} {026012} (\bibinfo {year} {2019})},\
  \Eprint {http://arxiv.org/abs/1905.03798} {arXiv:1905.03798 [hep-th]}
  \BibitemShut {NoStop}%
\bibitem [{\citenamefont {Groot~Nibbelink}\ and\ \citenamefont
  {Pospelov}(2005)}]{GrootNibbelink:2004za}%
  \BibitemOpen
  \bibfield  {author} {\bibinfo {author} {\bibfnamefont {S.}~\bibnamefont
  {Groot~Nibbelink}}\ and\ \bibinfo {author} {\bibfnamefont {M.}~\bibnamefont
  {Pospelov}},\ }\href {\doibase 10.1103/PhysRevLett.94.081601} {\bibfield
  {journal} {\bibinfo  {journal} {Phys. Rev. Lett.}\ }\textbf {\bibinfo
  {volume} {94}},\ \bibinfo {pages} {081601} (\bibinfo {year} {2005})},\
  \Eprint {http://arxiv.org/abs/hep-ph/0404271} {arXiv:hep-ph/0404271 [hep-ph]}
  \BibitemShut {NoStop}%
\bibitem [{\citenamefont {Pospelov}\ and\ \citenamefont
  {Shang}(2012)}]{Pospelov:2010mp}%
  \BibitemOpen
  \bibfield  {author} {\bibinfo {author} {\bibfnamefont {M.}~\bibnamefont
  {Pospelov}}\ and\ \bibinfo {author} {\bibfnamefont {Y.}~\bibnamefont
  {Shang}},\ }\href {\doibase 10.1103/PhysRevD.85.105001} {\bibfield  {journal}
  {\bibinfo  {journal} {Phys. Rev.}\ }\textbf {\bibinfo {volume} {D85}},\
  \bibinfo {pages} {105001} (\bibinfo {year} {2012})},\ \Eprint
  {http://arxiv.org/abs/1010.5249} {arXiv:1010.5249 [hep-th]} \BibitemShut
  {NoStop}%
\bibitem [{\citenamefont {Blas}\ \emph {et~al.}(2010)\citenamefont {Blas},
  \citenamefont {Pujolas},\ and\ \citenamefont {Sibiryakov}}]{Blas:2009qj}%
  \BibitemOpen
  \bibfield  {author} {\bibinfo {author} {\bibfnamefont {D.}~\bibnamefont
  {Blas}}, \bibinfo {author} {\bibfnamefont {O.}~\bibnamefont {Pujolas}}, \
  and\ \bibinfo {author} {\bibfnamefont {S.}~\bibnamefont {Sibiryakov}},\
  }\href {\doibase 10.1103/PhysRevLett.104.181302} {\bibfield  {journal}
  {\bibinfo  {journal} {Phys. Rev. Lett.}\ }\textbf {\bibinfo {volume} {104}},\
  \bibinfo {pages} {181302} (\bibinfo {year} {2010})},\ \Eprint
  {http://arxiv.org/abs/0909.3525} {arXiv:0909.3525 [hep-th]} \BibitemShut
  {NoStop}%
\bibitem [{\citenamefont {Blas}\ \emph {et~al.}(2011)\citenamefont {Blas},
  \citenamefont {Pujolas},\ and\ \citenamefont {Sibiryakov}}]{Blas:2010hb}%
  \BibitemOpen
  \bibfield  {author} {\bibinfo {author} {\bibfnamefont {D.}~\bibnamefont
  {Blas}}, \bibinfo {author} {\bibfnamefont {O.}~\bibnamefont {Pujolas}}, \
  and\ \bibinfo {author} {\bibfnamefont {S.}~\bibnamefont {Sibiryakov}},\
  }\href {\doibase 10.1007/JHEP04(2011)018} {\bibfield  {journal} {\bibinfo
  {journal} {JHEP}\ }\textbf {\bibinfo {volume} {04}},\ \bibinfo {pages} {018}
  (\bibinfo {year} {2011})},\ \Eprint {http://arxiv.org/abs/1007.3503}
  {arXiv:1007.3503 [hep-th]} \BibitemShut {NoStop}%
\bibitem [{\citenamefont {Jacobson}(2010)}]{Jacobson:2010mx}%
  \BibitemOpen
  \bibfield  {author} {\bibinfo {author} {\bibfnamefont {T.}~\bibnamefont
  {Jacobson}},\ }\href {\doibase 10.1103/PhysRevD.82.129901,
  10.1103/PhysRevD.81.101502} {\bibfield  {journal} {\bibinfo  {journal} {Phys.
  Rev.}\ }\textbf {\bibinfo {volume} {D81}},\ \bibinfo {pages} {101502}
  (\bibinfo {year} {2010})},\ \bibinfo {note} {[Erratum: Phys.
  Rev.D82,129901(2010)]},\ \Eprint {http://arxiv.org/abs/1001.4823}
  {arXiv:1001.4823 [hep-th]} \BibitemShut {NoStop}%
\bibitem [{\citenamefont {Abbott}\ \emph {et~al.}(2017)\citenamefont {Abbott}
  \emph {et~al.}}]{Monitor:2017mdv}%
  \BibitemOpen
  \bibfield  {author} {\bibinfo {author} {\bibfnamefont {B.~P.}\ \bibnamefont
  {Abbott}} \emph {et~al.} (\bibinfo {collaboration} {Virgo, Fermi-GBM,
  INTEGRAL, LIGO Scientific}),\ }\href {\doibase 10.3847/2041-8213/aa920c}
  {\bibfield  {journal} {\bibinfo  {journal} {Astrophys. J.}\ }\textbf
  {\bibinfo {volume} {848}},\ \bibinfo {pages} {L13} (\bibinfo {year}
  {2017})},\ \Eprint {http://arxiv.org/abs/1710.05834} {arXiv:1710.05834
  [astro-ph.HE]} \BibitemShut {NoStop}%
\bibitem [{\citenamefont {Emir~Gumrukcuoglu}\ \emph {et~al.}(2018)\citenamefont
  {Emir~Gumrukcuoglu}, \citenamefont {Saravani},\ and\ \citenamefont
  {Sotiriou}}]{Gumrukcuoglu:2017ijh}%
  \BibitemOpen
  \bibfield  {author} {\bibinfo {author} {\bibfnamefont {A.}~\bibnamefont
  {Emir~Gumrukcuoglu}}, \bibinfo {author} {\bibfnamefont {M.}~\bibnamefont
  {Saravani}}, \ and\ \bibinfo {author} {\bibfnamefont {T.~P.}\ \bibnamefont
  {Sotiriou}},\ }\href {\doibase 10.1103/PhysRevD.97.024032} {\bibfield
  {journal} {\bibinfo  {journal} {Phys. Rev.}\ }\textbf {\bibinfo {volume}
  {D97}},\ \bibinfo {pages} {024032} (\bibinfo {year} {2018})},\ \Eprint
  {http://arxiv.org/abs/1711.08845} {arXiv:1711.08845 [gr-qc]} \BibitemShut
  {NoStop}%
\bibitem [{\citenamefont {Will}(1993)}]{Will:1993hxu}%
  \BibitemOpen
  \bibfield  {author} {\bibinfo {author} {\bibfnamefont {C.~M.}\ \bibnamefont
  {Will}},\ }\href {\doibase 10.1017/CBO9780511564246} {\emph {\bibinfo {title}
  {Theory and Experiment in Gravitational Physics}}}\ (\bibinfo  {publisher}
  {Cambridge University Press},\ \bibinfo {year} {1993})\BibitemShut {NoStop}%
\bibitem [{\citenamefont {Will}(2014)}]{Will:2014kxa}%
  \BibitemOpen
  \bibfield  {author} {\bibinfo {author} {\bibfnamefont {C.~M.}\ \bibnamefont
  {Will}},\ }\href {\doibase 10.12942/lrr-2014-4} {\bibfield  {journal}
  {\bibinfo  {journal} {Living Rev. Rel.}\ }\textbf {\bibinfo {volume} {17}},\
  \bibinfo {pages} {4} (\bibinfo {year} {2014})},\ \Eprint
  {http://arxiv.org/abs/1403.7377} {arXiv:1403.7377 [gr-qc]} \BibitemShut
  {NoStop}%
\bibitem [{\citenamefont {Blas}\ and\ \citenamefont
  {Sanctuary}(2011)}]{Blas:2011zd}%
  \BibitemOpen
  \bibfield  {author} {\bibinfo {author} {\bibfnamefont {D.}~\bibnamefont
  {Blas}}\ and\ \bibinfo {author} {\bibfnamefont {H.}~\bibnamefont
  {Sanctuary}},\ }\href {\doibase 10.1103/PhysRevD.84.064004} {\bibfield
  {journal} {\bibinfo  {journal} {Phys. Rev.}\ }\textbf {\bibinfo {volume}
  {D84}},\ \bibinfo {pages} {064004} (\bibinfo {year} {2011})},\ \Eprint
  {http://arxiv.org/abs/1105.5149} {arXiv:1105.5149 [gr-qc]} \BibitemShut
  {NoStop}%
\bibitem [{\citenamefont {Bonetti}\ and\ \citenamefont
  {Barausse}(2015)}]{Bonetti:2015oda}%
  \BibitemOpen
  \bibfield  {author} {\bibinfo {author} {\bibfnamefont {M.}~\bibnamefont
  {Bonetti}}\ and\ \bibinfo {author} {\bibfnamefont {E.}~\bibnamefont
  {Barausse}},\ }\href {\doibase 10.1103/PhysRevD.91.084053,
  10.1103/PhysRevD.93.029901} {\bibfield  {journal} {\bibinfo  {journal} {Phys.
  Rev.}\ }\textbf {\bibinfo {volume} {D91}},\ \bibinfo {pages} {084053}
  (\bibinfo {year} {2015})},\ \bibinfo {note} {[Erratum: Phys.
  Rev.D93,029901(2016)]},\ \Eprint {http://arxiv.org/abs/1502.05554}
  {arXiv:1502.05554 [gr-qc]} \BibitemShut {NoStop}%
\bibitem [{\citenamefont {Ramos}\ and\ \citenamefont {Barausse}(2019)}]{ramos}%
  \BibitemOpen
  \bibfield  {author} {\bibinfo {author} {\bibfnamefont {O.}~\bibnamefont
  {Ramos}}\ and\ \bibinfo {author} {\bibfnamefont {E.}~\bibnamefont
  {Barausse}},\ }\href {\doibase 10.1103/PhysRevD.99.024034} {\bibfield
  {journal} {\bibinfo  {journal} {Phys. Rev.}\ }\textbf {\bibinfo {volume}
  {D99}},\ \bibinfo {pages} {024034} (\bibinfo {year} {2019})},\ \Eprint
  {http://arxiv.org/abs/1811.07786} {arXiv:1811.07786 [gr-qc]} \BibitemShut
  {NoStop}%
\bibitem [{\citenamefont {Carroll}\ and\ \citenamefont
  {Lim}(2004)}]{Carroll:2004ai}%
  \BibitemOpen
  \bibfield  {author} {\bibinfo {author} {\bibfnamefont {S.~M.}\ \bibnamefont
  {Carroll}}\ and\ \bibinfo {author} {\bibfnamefont {E.~A.}\ \bibnamefont
  {Lim}},\ }\href {\doibase 10.1103/PhysRevD.70.123525} {\bibfield  {journal}
  {\bibinfo  {journal} {Phys. Rev.}\ }\textbf {\bibinfo {volume} {D70}},\
  \bibinfo {pages} {123525} (\bibinfo {year} {2004})},\ \Eprint
  {http://arxiv.org/abs/hep-th/0407149} {arXiv:hep-th/0407149 [hep-th]}
  \BibitemShut {NoStop}%
\bibitem [{\citenamefont {Yagi}\ \emph
  {et~al.}(2014{\natexlab{a}})\citenamefont {Yagi}, \citenamefont {Blas},
  \citenamefont {Barausse},\ and\ \citenamefont {Yunes}}]{Yagi:2013ava}%
  \BibitemOpen
  \bibfield  {author} {\bibinfo {author} {\bibfnamefont {K.}~\bibnamefont
  {Yagi}}, \bibinfo {author} {\bibfnamefont {D.}~\bibnamefont {Blas}}, \bibinfo
  {author} {\bibfnamefont {E.}~\bibnamefont {Barausse}}, \ and\ \bibinfo
  {author} {\bibfnamefont {N.}~\bibnamefont {Yunes}},\ }\href {\doibase
  10.1103/PhysRevD.90.069902, 10.1103/PhysRevD.90.069901,
  10.1103/PhysRevD.89.084067} {\bibfield  {journal} {\bibinfo  {journal} {Phys.
  Rev.}\ }\textbf {\bibinfo {volume} {D89}},\ \bibinfo {pages} {084067}
  (\bibinfo {year} {2014}{\natexlab{a}})},\ \bibinfo {note} {[Erratum: Phys.
  Rev.D90,no.6,069901(2014)]},\ \Eprint {http://arxiv.org/abs/1311.7144}
  {arXiv:1311.7144 [gr-qc]} \BibitemShut {NoStop}%
\bibitem [{\citenamefont {Yagi}\ \emph
  {et~al.}(2014{\natexlab{b}})\citenamefont {Yagi}, \citenamefont {Blas},
  \citenamefont {Yunes},\ and\ \citenamefont {Barausse}}]{Yagi:2013qpa}%
  \BibitemOpen
  \bibfield  {author} {\bibinfo {author} {\bibfnamefont {K.}~\bibnamefont
  {Yagi}}, \bibinfo {author} {\bibfnamefont {D.}~\bibnamefont {Blas}}, \bibinfo
  {author} {\bibfnamefont {N.}~\bibnamefont {Yunes}}, \ and\ \bibinfo {author}
  {\bibfnamefont {E.}~\bibnamefont {Barausse}},\ }\href {\doibase
  10.1103/PhysRevLett.112.161101} {\bibfield  {journal} {\bibinfo  {journal}
  {Phys. Rev. Lett.}\ }\textbf {\bibinfo {volume} {112}},\ \bibinfo {pages}
  {161101} (\bibinfo {year} {2014}{\natexlab{b}})},\ \Eprint
  {http://arxiv.org/abs/1307.6219} {arXiv:1307.6219 [gr-qc]} \BibitemShut
  {NoStop}%
\bibitem [{\citenamefont {Berglund}\ \emph {et~al.}(2012)\citenamefont
  {Berglund}, \citenamefont {Bhattacharyya},\ and\ \citenamefont
  {Mattingly}}]{Berglund:2012bu}%
  \BibitemOpen
  \bibfield  {author} {\bibinfo {author} {\bibfnamefont {P.}~\bibnamefont
  {Berglund}}, \bibinfo {author} {\bibfnamefont {J.}~\bibnamefont
  {Bhattacharyya}}, \ and\ \bibinfo {author} {\bibfnamefont {D.}~\bibnamefont
  {Mattingly}},\ }\href {\doibase 10.1103/PhysRevD.85.124019} {\bibfield
  {journal} {\bibinfo  {journal} {Phys. Rev.}\ }\textbf {\bibinfo {volume}
  {D85}},\ \bibinfo {pages} {124019} (\bibinfo {year} {2012})},\ \Eprint
  {http://arxiv.org/abs/1202.4497} {arXiv:1202.4497 [hep-th]} \BibitemShut
  {NoStop}%
\bibitem [{\citenamefont {Barausse}(2019)}]{Barausse:2019yuk}%
  \BibitemOpen
  \bibfield  {author} {\bibinfo {author} {\bibfnamefont {E.}~\bibnamefont
  {Barausse}},\ }\href {\doibase 10.1103/PhysRevD.100.084053} {\bibfield
  {journal} {\bibinfo  {journal} {Phys. Rev. D}\ }\textbf {\bibinfo {volume}
  {100}},\ \bibinfo {pages} {084053} (\bibinfo {year} {2019})},\ \Eprint
  {http://arxiv.org/abs/1907.05958} {arXiv:1907.05958 [gr-qc]} \BibitemShut
  {NoStop}%
\bibitem [{\citenamefont {Will}(2018)}]{Will:2018ont}%
  \BibitemOpen
  \bibfield  {author} {\bibinfo {author} {\bibfnamefont {C.~M.}\ \bibnamefont
  {Will}},\ }\href {\doibase 10.1088/1361-6382/aab1c6} {\bibfield  {journal}
  {\bibinfo  {journal} {Class. Quant. Grav.}\ }\textbf {\bibinfo {volume}
  {35}},\ \bibinfo {pages} {085001} (\bibinfo {year} {2018})},\ \Eprint
  {http://arxiv.org/abs/1801.08999} {arXiv:1801.08999 [gr-qc]} \BibitemShut
  {NoStop}%
\bibitem [{\citenamefont {Kobayashi}\ \emph {et~al.}(2010)\citenamefont
  {Kobayashi}, \citenamefont {Urakawa},\ and\ \citenamefont
  {Yamaguchi}}]{Kobayashi:2010eh}%
  \BibitemOpen
  \bibfield  {author} {\bibinfo {author} {\bibfnamefont {T.}~\bibnamefont
  {Kobayashi}}, \bibinfo {author} {\bibfnamefont {Y.}~\bibnamefont {Urakawa}},
  \ and\ \bibinfo {author} {\bibfnamefont {M.}~\bibnamefont {Yamaguchi}},\
  }\href {\doibase 10.1088/1475-7516/2010/04/025} {\bibfield  {journal}
  {\bibinfo  {journal} {JCAP}\ }\textbf {\bibinfo {volume} {04}},\ \bibinfo
  {pages} {025} (\bibinfo {year} {2010})},\ \Eprint
  {http://arxiv.org/abs/1002.3101} {arXiv:1002.3101 [hep-th]} \BibitemShut
  {NoStop}%
\bibitem [{\citenamefont {Barausse}\ and\ \citenamefont
  {Lehner}(2013)}]{Barausse:2013ysa}%
  \BibitemOpen
  \bibfield  {author} {\bibinfo {author} {\bibfnamefont {E.}~\bibnamefont
  {Barausse}}\ and\ \bibinfo {author} {\bibfnamefont {L.}~\bibnamefont
  {Lehner}},\ }\href {\doibase 10.1103/PhysRevD.88.024029} {\bibfield
  {journal} {\bibinfo  {journal} {Phys. Rev. D}\ }\textbf {\bibinfo {volume}
  {88}},\ \bibinfo {pages} {024029} (\bibinfo {year} {2013})},\ \Eprint
  {http://arxiv.org/abs/1306.5564} {arXiv:1306.5564 [gr-qc]} \BibitemShut
  {NoStop}%
\bibitem [{\citenamefont {{Loll}}\ and\ \citenamefont
  {{Pires}}(2014)}]{2014arXiv1407.1259L}%
  \BibitemOpen
  \bibfield  {author} {\bibinfo {author} {\bibfnamefont {R.}~\bibnamefont
  {{Loll}}}\ and\ \bibinfo {author} {\bibfnamefont {L.}~\bibnamefont
  {{Pires}}},\ }\href@noop {} {\bibfield  {journal} {\bibinfo  {journal} {ArXiv
  e-prints}\ } (\bibinfo {year} {2014})},\ \Eprint
  {http://arxiv.org/abs/1407.1259} {arXiv:1407.1259 [hep-th]} \BibitemShut
  {NoStop}%
\bibitem [{\citenamefont {Bellorin}\ and\ \citenamefont
  {Restuccia}(2012)}]{Bellorin:2010je}%
  \BibitemOpen
  \bibfield  {author} {\bibinfo {author} {\bibfnamefont {J.}~\bibnamefont
  {Bellorin}}\ and\ \bibinfo {author} {\bibfnamefont {A.}~\bibnamefont
  {Restuccia}},\ }\href {\doibase 10.1142/S021827182500290} {\bibfield
  {journal} {\bibinfo  {journal} {Int. J. Mod. Phys.}\ }\textbf {\bibinfo
  {volume} {D21}},\ \bibinfo {pages} {1250029} (\bibinfo {year} {2012})},\
  \Eprint {http://arxiv.org/abs/1004.0055} {arXiv:1004.0055 [hep-th]}
  \BibitemShut {NoStop}%
\bibitem [{\citenamefont {Henneaux}\ \emph {et~al.}(2010)\citenamefont
  {Henneaux}, \citenamefont {Kleinschmidt},\ and\ \citenamefont
  {Lucena~Gómez}}]{Henneaux:2009zb}%
  \BibitemOpen
  \bibfield  {author} {\bibinfo {author} {\bibfnamefont {M.}~\bibnamefont
  {Henneaux}}, \bibinfo {author} {\bibfnamefont {A.}~\bibnamefont
  {Kleinschmidt}}, \ and\ \bibinfo {author} {\bibfnamefont {G.}~\bibnamefont
  {Lucena~Gómez}},\ }\href {\doibase 10.1103/PhysRevD.81.064002} {\bibfield
  {journal} {\bibinfo  {journal} {Phys. Rev. D}\ }\textbf {\bibinfo {volume}
  {81}},\ \bibinfo {pages} {064002} (\bibinfo {year} {2010})},\ \Eprint
  {http://arxiv.org/abs/0912.0399} {arXiv:0912.0399 [hep-th]} \BibitemShut
  {NoStop}%
\bibitem [{\citenamefont {Barausse}\ \emph {et~al.}(2011)\citenamefont
  {Barausse}, \citenamefont {Jacobson},\ and\ \citenamefont
  {Sotiriou}}]{Barausse:2011pu}%
  \BibitemOpen
  \bibfield  {author} {\bibinfo {author} {\bibfnamefont {E.}~\bibnamefont
  {Barausse}}, \bibinfo {author} {\bibfnamefont {T.}~\bibnamefont {Jacobson}},
  \ and\ \bibinfo {author} {\bibfnamefont {T.~P.}\ \bibnamefont {Sotiriou}},\
  }\href {\doibase 10.1103/PhysRevD.83.124043} {\bibfield  {journal} {\bibinfo
  {journal} {Phys. Rev.}\ }\textbf {\bibinfo {volume} {D83}},\ \bibinfo {pages}
  {124043} (\bibinfo {year} {2011})},\ \Eprint {http://arxiv.org/abs/1104.2889}
  {arXiv:1104.2889 [gr-qc]} \BibitemShut {NoStop}%
\bibitem [{\citenamefont {Blas}\ and\ \citenamefont
  {Sibiryakov}(2011)}]{Blas:2011ni}%
  \BibitemOpen
  \bibfield  {author} {\bibinfo {author} {\bibfnamefont {D.}~\bibnamefont
  {Blas}}\ and\ \bibinfo {author} {\bibfnamefont {S.}~\bibnamefont
  {Sibiryakov}},\ }\href {\doibase 10.1103/PhysRevD.84.124043} {\bibfield
  {journal} {\bibinfo  {journal} {Phys. Rev.}\ }\textbf {\bibinfo {volume}
  {D84}},\ \bibinfo {pages} {124043} (\bibinfo {year} {2011})},\ \Eprint
  {http://arxiv.org/abs/1110.2195} {arXiv:1110.2195 [hep-th]} \BibitemShut
  {NoStop}%
\bibitem [{\citenamefont {Jacobson}\ and\ \citenamefont
  {Mattingly}(2004)}]{Jacobson:2004ts}%
  \BibitemOpen
  \bibfield  {author} {\bibinfo {author} {\bibfnamefont {T.}~\bibnamefont
  {Jacobson}}\ and\ \bibinfo {author} {\bibfnamefont {D.}~\bibnamefont
  {Mattingly}},\ }\href {\doibase 10.1103/PhysRevD.70.024003} {\bibfield
  {journal} {\bibinfo  {journal} {Phys. Rev.}\ }\textbf {\bibinfo {volume}
  {D70}},\ \bibinfo {pages} {024003} (\bibinfo {year} {2004})},\ \Eprint
  {http://arxiv.org/abs/gr-qc/0402005} {arXiv:gr-qc/0402005 [gr-qc]}
  \BibitemShut {NoStop}%
\bibitem [{\citenamefont {Elliott}\ \emph {et~al.}(2005)\citenamefont
  {Elliott}, \citenamefont {Moore},\ and\ \citenamefont
  {Stoica}}]{Elliott:2005va}%
  \BibitemOpen
  \bibfield  {author} {\bibinfo {author} {\bibfnamefont {J.~W.}\ \bibnamefont
  {Elliott}}, \bibinfo {author} {\bibfnamefont {G.~D.}\ \bibnamefont {Moore}},
  \ and\ \bibinfo {author} {\bibfnamefont {H.}~\bibnamefont {Stoica}},\ }\href
  {\doibase 10.1088/1126-6708/2005/08/066} {\bibfield  {journal} {\bibinfo
  {journal} {JHEP}\ }\textbf {\bibinfo {volume} {08}},\ \bibinfo {pages} {066}
  (\bibinfo {year} {2005})},\ \Eprint {http://arxiv.org/abs/hep-ph/0505211}
  {arXiv:hep-ph/0505211 [hep-ph]} \BibitemShut {NoStop}%
\bibitem [{\citenamefont {Audren}\ \emph {et~al.}(2013)\citenamefont {Audren},
  \citenamefont {Blas}, \citenamefont {Lesgourgues},\ and\ \citenamefont
  {Sibiryakov}}]{Audren:2013dwa}%
  \BibitemOpen
  \bibfield  {author} {\bibinfo {author} {\bibfnamefont {B.}~\bibnamefont
  {Audren}}, \bibinfo {author} {\bibfnamefont {D.}~\bibnamefont {Blas}},
  \bibinfo {author} {\bibfnamefont {J.}~\bibnamefont {Lesgourgues}}, \ and\
  \bibinfo {author} {\bibfnamefont {S.}~\bibnamefont {Sibiryakov}},\ }\href
  {\doibase 10.1088/1475-7516/2013/08/039} {\bibfield  {journal} {\bibinfo
  {journal} {JCAP}\ }\textbf {\bibinfo {volume} {1308}},\ \bibinfo {pages}
  {039} (\bibinfo {year} {2013})},\ \Eprint {http://arxiv.org/abs/1305.0009}
  {arXiv:1305.0009 [astro-ph.CO]} \BibitemShut {NoStop}%
\bibitem [{\citenamefont {Eling}\ and\ \citenamefont
  {Jacobson}(2006{\natexlab{a}})}]{Eling:2006ec}%
  \BibitemOpen
  \bibfield  {author} {\bibinfo {author} {\bibfnamefont {C.}~\bibnamefont
  {Eling}}\ and\ \bibinfo {author} {\bibfnamefont {T.}~\bibnamefont
  {Jacobson}},\ }\href {\doibase 10.1088/0264-9381/23/18/009,
  10.1088/0264-9381/27/4/049802} {\bibfield  {journal} {\bibinfo  {journal}
  {Class. Quant. Grav.}\ }\textbf {\bibinfo {volume} {23}},\ \bibinfo {pages}
  {5643} (\bibinfo {year} {2006}{\natexlab{a}})},\ \bibinfo {note} {[Erratum:
  Class. Quant. Grav.27,049802(2010)]},\ \Eprint
  {http://arxiv.org/abs/gr-qc/0604088} {arXiv:gr-qc/0604088 [gr-qc]}
  \BibitemShut {NoStop}%
\bibitem [{\citenamefont {Eling}\ and\ \citenamefont
  {Jacobson}(2006{\natexlab{b}})}]{Eling:2006df}%
  \BibitemOpen
  \bibfield  {author} {\bibinfo {author} {\bibfnamefont {C.}~\bibnamefont
  {Eling}}\ and\ \bibinfo {author} {\bibfnamefont {T.}~\bibnamefont
  {Jacobson}},\ }\href {\doibase 10.1088/0264-9381/23/18/008} {\bibfield
  {journal} {\bibinfo  {journal} {Class. Quant. Grav.}\ }\textbf {\bibinfo
  {volume} {23}},\ \bibinfo {pages} {5625} (\bibinfo {year}
  {2006}{\natexlab{b}})},\ \bibinfo {note} {[Erratum: Class.Quant.Grav. 27,
  049801 (2010)]},\ \Eprint {http://arxiv.org/abs/gr-qc/0603058}
  {arXiv:gr-qc/0603058} \BibitemShut {NoStop}%
\bibitem [{\citenamefont {Eling}\ \emph {et~al.}(2007)\citenamefont {Eling},
  \citenamefont {Jacobson},\ and\ \citenamefont
  {Coleman~Miller}}]{Eling:2007xh}%
  \BibitemOpen
  \bibfield  {author} {\bibinfo {author} {\bibfnamefont {C.}~\bibnamefont
  {Eling}}, \bibinfo {author} {\bibfnamefont {T.}~\bibnamefont {Jacobson}}, \
  and\ \bibinfo {author} {\bibfnamefont {M.}~\bibnamefont {Coleman~Miller}},\
  }\href {\doibase 10.1103/PhysRevD.76.042003, 10.1103/PhysRevD.80.129906}
  {\bibfield  {journal} {\bibinfo  {journal} {Phys. Rev.}\ }\textbf {\bibinfo
  {volume} {D76}},\ \bibinfo {pages} {042003} (\bibinfo {year} {2007})},\
  \bibinfo {note} {[Erratum: Phys. Rev.D80,129906(2009)]},\ \Eprint
  {http://arxiv.org/abs/0705.1565} {arXiv:0705.1565 [gr-qc]} \BibitemShut
  {NoStop}%
\bibitem [{\citenamefont {Garfinkle}\ \emph {et~al.}(2007)\citenamefont
  {Garfinkle}, \citenamefont {Eling},\ and\ \citenamefont
  {Jacobson}}]{Garfinkle:2007bk}%
  \BibitemOpen
  \bibfield  {author} {\bibinfo {author} {\bibfnamefont {D.}~\bibnamefont
  {Garfinkle}}, \bibinfo {author} {\bibfnamefont {C.}~\bibnamefont {Eling}}, \
  and\ \bibinfo {author} {\bibfnamefont {T.}~\bibnamefont {Jacobson}},\ }\href
  {\doibase 10.1103/PhysRevD.76.024003} {\bibfield  {journal} {\bibinfo
  {journal} {Phys. Rev.}\ }\textbf {\bibinfo {volume} {D76}},\ \bibinfo {pages}
  {024003} (\bibinfo {year} {2007})},\ \Eprint
  {http://arxiv.org/abs/gr-qc/0703093} {arXiv:gr-qc/0703093 [GR-QC]}
  \BibitemShut {NoStop}%
\bibitem [{\citenamefont {Estabrook}\ \emph {et~al.}(1973)\citenamefont
  {Estabrook}, \citenamefont {Wahlquist}, \citenamefont {Christensen},
  \citenamefont {DeWitt}, \citenamefont {Smarr},\ and\ \citenamefont
  {Tsiang}}]{Estabrook:1973ue}%
  \BibitemOpen
  \bibfield  {author} {\bibinfo {author} {\bibfnamefont {F.}~\bibnamefont
  {Estabrook}}, \bibinfo {author} {\bibfnamefont {H.}~\bibnamefont
  {Wahlquist}}, \bibinfo {author} {\bibfnamefont {S.}~\bibnamefont
  {Christensen}}, \bibinfo {author} {\bibfnamefont {B.}~\bibnamefont {DeWitt}},
  \bibinfo {author} {\bibfnamefont {L.}~\bibnamefont {Smarr}}, \ and\ \bibinfo
  {author} {\bibfnamefont {E.}~\bibnamefont {Tsiang}},\ }\href {\doibase
  10.1103/PhysRevD.7.2814} {\bibfield  {journal} {\bibinfo  {journal} {Phys.
  Rev. D}\ }\textbf {\bibinfo {volume} {7}},\ \bibinfo {pages} {2814} (\bibinfo
  {year} {1973})}\BibitemShut {NoStop}%
\bibitem [{\citenamefont {Petrich}\ \emph {et~al.}(1985)\citenamefont
  {Petrich}, \citenamefont {Shapiro},\ and\ \citenamefont
  {Teukolsky}}]{Petrich:1985jko}%
  \BibitemOpen
  \bibfield  {author} {\bibinfo {author} {\bibfnamefont {L.~I.}\ \bibnamefont
  {Petrich}}, \bibinfo {author} {\bibfnamefont {S.~L.}\ \bibnamefont
  {Shapiro}}, \ and\ \bibinfo {author} {\bibfnamefont {S.~A.}\ \bibnamefont
  {Teukolsky}},\ }\href {\doibase 10.1103/PhysRevD.31.2459} {\bibfield
  {journal} {\bibinfo  {journal} {Phys. Rev. D}\ }\textbf {\bibinfo {volume}
  {31}},\ \bibinfo {pages} {2459} (\bibinfo {year} {1985})}\BibitemShut
  {NoStop}%
\bibitem [{\citenamefont {{Shapiro}}\ and\ \citenamefont
  {{Teukolsky}}(1985)}]{1985ApJ...298...34S}%
  \BibitemOpen
  \bibfield  {author} {\bibinfo {author} {\bibfnamefont {S.~L.}\ \bibnamefont
  {{Shapiro}}}\ and\ \bibinfo {author} {\bibfnamefont {S.~A.}\ \bibnamefont
  {{Teukolsky}}},\ }\href {\doibase 10.1086/163587} {\bibfield  {journal}
  {\bibinfo  {journal} {\apj}\ }\textbf {\bibinfo {volume} {298}},\ \bibinfo
  {pages} {34} (\bibinfo {year} {1985})}\BibitemShut {NoStop}%
\bibitem [{\citenamefont {Beig}\ and\ \citenamefont
  {O'Murchadha}(1998)}]{Beig:1997fp}%
  \BibitemOpen
  \bibfield  {author} {\bibinfo {author} {\bibfnamefont {R.}~\bibnamefont
  {Beig}}\ and\ \bibinfo {author} {\bibfnamefont {N.}~\bibnamefont
  {O'Murchadha}},\ }\href {\doibase 10.1103/PhysRevD.57.4728} {\bibfield
  {journal} {\bibinfo  {journal} {Phys. Rev. D}\ }\textbf {\bibinfo {volume}
  {57}},\ \bibinfo {pages} {4728} (\bibinfo {year} {1998})},\ \Eprint
  {http://arxiv.org/abs/gr-qc/9706046} {arXiv:gr-qc/9706046} \BibitemShut
  {NoStop}%
\bibitem [{\citenamefont {Alcubierre}(2008)}]{Alcubierre:1138167}%
  \BibitemOpen
  \bibfield  {author} {\bibinfo {author} {\bibfnamefont {M.}~\bibnamefont
  {Alcubierre}},\ }\href {\doibase 10.1093/acprof:oso/9780199205677.001.0001}
  {\emph {\bibinfo {title} {{Introduction to 3+1 numerical relativity}}}},\
  International series of monographs on physics\ (\bibinfo  {publisher} {Oxford
  Univ. Press},\ \bibinfo {address} {Oxford},\ \bibinfo {year}
  {2008})\BibitemShut {NoStop}%
\bibitem [{\citenamefont {Baumgarte}\ and\ \citenamefont
  {Shapiro}(2010)}]{Baumgarte:2010ndz}%
  \BibitemOpen
  \bibfield  {author} {\bibinfo {author} {\bibfnamefont {T.~W.}\ \bibnamefont
  {Baumgarte}}\ and\ \bibinfo {author} {\bibfnamefont {S.~L.}\ \bibnamefont
  {Shapiro}},\ }\href {\doibase 10.1017/CBO9781139193344} {\emph {\bibinfo
  {title} {{Numerical Relativity: Solving Einstein's Equations on the
  Computer}}}}\ (\bibinfo  {publisher} {Cambridge University Press},\ \bibinfo
  {year} {2010})\BibitemShut {NoStop}%
\bibitem [{\citenamefont {Loll}\ and\ \citenamefont
  {Pires}(2014)}]{Loll:2014xja}%
  \BibitemOpen
  \bibfield  {author} {\bibinfo {author} {\bibfnamefont {R.}~\bibnamefont
  {Loll}}\ and\ \bibinfo {author} {\bibfnamefont {L.}~\bibnamefont {Pires}},\
  }\href {\doibase 10.1103/PhysRevD.90.124050} {\bibfield  {journal} {\bibinfo
  {journal} {Phys. Rev. D}\ }\textbf {\bibinfo {volume} {90}},\ \bibinfo
  {pages} {124050} (\bibinfo {year} {2014})},\ \Eprint
  {http://arxiv.org/abs/1407.1259} {arXiv:1407.1259 [hep-th]} \BibitemShut
  {NoStop}%
\bibitem [{\citenamefont {Markovic}\ and\ \citenamefont
  {Shapiro}(2000)}]{Markovic:1999di}%
  \BibitemOpen
  \bibfield  {author} {\bibinfo {author} {\bibfnamefont {D.}~\bibnamefont
  {Markovic}}\ and\ \bibinfo {author} {\bibfnamefont {S.}~\bibnamefont
  {Shapiro}},\ }\href {\doibase 10.1103/PhysRevD.61.084029} {\bibfield
  {journal} {\bibinfo  {journal} {Phys. Rev. D}\ }\textbf {\bibinfo {volume}
  {61}},\ \bibinfo {pages} {084029} (\bibinfo {year} {2000})},\ \Eprint
  {http://arxiv.org/abs/gr-qc/9912066} {arXiv:gr-qc/9912066} \BibitemShut
  {NoStop}%
\bibitem [{\citenamefont {Saravani}\ \emph {et~al.}(2014)\citenamefont
  {Saravani}, \citenamefont {Afshordi},\ and\ \citenamefont
  {Mann}}]{Saravani:2013kva}%
  \BibitemOpen
  \bibfield  {author} {\bibinfo {author} {\bibfnamefont {M.}~\bibnamefont
  {Saravani}}, \bibinfo {author} {\bibfnamefont {N.}~\bibnamefont {Afshordi}},
  \ and\ \bibinfo {author} {\bibfnamefont {R.~B.}\ \bibnamefont {Mann}},\
  }\href {\doibase 10.1103/PhysRevD.89.084029} {\bibfield  {journal} {\bibinfo
  {journal} {Phys. Rev. D}\ }\textbf {\bibinfo {volume} {89}},\ \bibinfo
  {pages} {084029} (\bibinfo {year} {2014})},\ \Eprint
  {http://arxiv.org/abs/1310.4143} {arXiv:1310.4143 [gr-qc]} \BibitemShut
  {NoStop}%
\bibitem [{\citenamefont {Afshordi}(2009)}]{Afshordi:2009tt}%
  \BibitemOpen
  \bibfield  {author} {\bibinfo {author} {\bibfnamefont {N.}~\bibnamefont
  {Afshordi}},\ }\href {\doibase 10.1103/PhysRevD.80.081502} {\bibfield
  {journal} {\bibinfo  {journal} {Phys. Rev.}\ }\textbf {\bibinfo {volume}
  {D80}},\ \bibinfo {pages} {081502} (\bibinfo {year} {2009})},\ \Eprint
  {http://arxiv.org/abs/0907.5201} {arXiv:0907.5201 [hep-th]} \BibitemShut
  {NoStop}%
\bibitem [{\citenamefont {Bhattacharyya}\ \emph {et~al.}(2018)\citenamefont
  {Bhattacharyya}, \citenamefont {Coates}, \citenamefont {Colombo},
  \citenamefont {Gumrukcuoglu},\ and\ \citenamefont
  {Sotiriou}}]{Bhattacharyya:2016mah}%
  \BibitemOpen
  \bibfield  {author} {\bibinfo {author} {\bibfnamefont {J.}~\bibnamefont
  {Bhattacharyya}}, \bibinfo {author} {\bibfnamefont {A.}~\bibnamefont
  {Coates}}, \bibinfo {author} {\bibfnamefont {M.}~\bibnamefont {Colombo}},
  \bibinfo {author} {\bibfnamefont {A.~E.}\ \bibnamefont {Gumrukcuoglu}}, \
  and\ \bibinfo {author} {\bibfnamefont {T.~P.}\ \bibnamefont {Sotiriou}},\
  }\href {\doibase 10.1103/PhysRevD.97.064020} {\bibfield  {journal} {\bibinfo
  {journal} {Phys. Rev. D}\ }\textbf {\bibinfo {volume} {97}},\ \bibinfo
  {pages} {064020} (\bibinfo {year} {2018})},\ \Eprint
  {http://arxiv.org/abs/1612.01824} {arXiv:1612.01824 [hep-th]} \BibitemShut
  {NoStop}%
\bibitem [{\citenamefont {Regge}\ and\ \citenamefont
  {Wheeler}(1957)}]{Regge:1957td}%
  \BibitemOpen
  \bibfield  {author} {\bibinfo {author} {\bibfnamefont {T.}~\bibnamefont
  {Regge}}\ and\ \bibinfo {author} {\bibfnamefont {J.~A.}\ \bibnamefont
  {Wheeler}},\ }\href {\doibase 10.1103/PhysRev.108.1063} {\bibfield  {journal}
  {\bibinfo  {journal} {Phys. Rev.}\ }\textbf {\bibinfo {volume} {108}},\
  \bibinfo {pages} {1063} (\bibinfo {year} {1957})}\BibitemShut {NoStop}%
\bibitem [{\citenamefont {Zerilli}(1970)}]{Zerilli:1970se}%
  \BibitemOpen
  \bibfield  {author} {\bibinfo {author} {\bibfnamefont {F.~J.}\ \bibnamefont
  {Zerilli}},\ }\href {\doibase 10.1103/PhysRevLett.24.737} {\bibfield
  {journal} {\bibinfo  {journal} {Phys. Rev. Lett.}\ }\textbf {\bibinfo
  {volume} {24}},\ \bibinfo {pages} {737} (\bibinfo {year} {1970})}\BibitemShut
  {NoStop}%
\bibitem [{\citenamefont {Teukolsky}(1973)}]{Teukolsky:1973ha}%
  \BibitemOpen
  \bibfield  {author} {\bibinfo {author} {\bibfnamefont {S.~A.}\ \bibnamefont
  {Teukolsky}},\ }\href {\doibase 10.1086/152444} {\bibfield  {journal}
  {\bibinfo  {journal} {Astrophys. J.}\ }\textbf {\bibinfo {volume} {185}},\
  \bibinfo {pages} {635} (\bibinfo {year} {1973})}\BibitemShut {NoStop}%
\bibitem [{\citenamefont {Israel}(1967)}]{Israel:1967wq}%
  \BibitemOpen
  \bibfield  {author} {\bibinfo {author} {\bibfnamefont {W.}~\bibnamefont
  {Israel}},\ }\href {\doibase 10.1103/PhysRev.164.1776} {\bibfield  {journal}
  {\bibinfo  {journal} {Phys. Rev.}\ }\textbf {\bibinfo {volume} {164}},\
  \bibinfo {pages} {1776} (\bibinfo {year} {1967})}\BibitemShut {NoStop}%
\bibitem [{\citenamefont {Hawking}(1972)}]{Hawking:1971vc}%
  \BibitemOpen
  \bibfield  {author} {\bibinfo {author} {\bibfnamefont {S.~W.}\ \bibnamefont
  {Hawking}},\ }\href {\doibase 10.1007/BF01877517} {\bibfield  {journal}
  {\bibinfo  {journal} {Commun. Math. Phys.}\ }\textbf {\bibinfo {volume}
  {25}},\ \bibinfo {pages} {152} (\bibinfo {year} {1972})}\BibitemShut
  {NoStop}%
\bibitem [{\citenamefont {Carter}(1971)}]{Carter:1971zc}%
  \BibitemOpen
  \bibfield  {author} {\bibinfo {author} {\bibfnamefont {B.}~\bibnamefont
  {Carter}},\ }\href {\doibase 10.1103/PhysRevLett.26.331} {\bibfield
  {journal} {\bibinfo  {journal} {Phys. Rev. Lett.}\ }\textbf {\bibinfo
  {volume} {26}},\ \bibinfo {pages} {331} (\bibinfo {year} {1971})}\BibitemShut
  {NoStop}%
\bibitem [{\citenamefont {Robinson}(1975)}]{Robinson:1975bv}%
  \BibitemOpen
  \bibfield  {author} {\bibinfo {author} {\bibfnamefont {D.}~\bibnamefont
  {Robinson}},\ }\href {\doibase 10.1103/PhysRevLett.34.905} {\bibfield
  {journal} {\bibinfo  {journal} {Phys. Rev. Lett.}\ }\textbf {\bibinfo
  {volume} {34}},\ \bibinfo {pages} {905} (\bibinfo {year} {1975})}\BibitemShut
  {NoStop}%
\bibitem [{\citenamefont {Barausse}\ \emph {et~al.}(2014)\citenamefont
  {Barausse}, \citenamefont {Cardoso},\ and\ \citenamefont
  {Pani}}]{Barausse:2014tra}%
  \BibitemOpen
  \bibfield  {author} {\bibinfo {author} {\bibfnamefont {E.}~\bibnamefont
  {Barausse}}, \bibinfo {author} {\bibfnamefont {V.}~\bibnamefont {Cardoso}}, \
  and\ \bibinfo {author} {\bibfnamefont {P.}~\bibnamefont {Pani}},\ }\href
  {\doibase 10.1103/PhysRevD.89.104059} {\bibfield  {journal} {\bibinfo
  {journal} {Phys. Rev.}\ }\textbf {\bibinfo {volume} {D89}},\ \bibinfo {pages}
  {104059} (\bibinfo {year} {2014})},\ \Eprint {http://arxiv.org/abs/1404.7149}
  {arXiv:1404.7149 [gr-qc]} \BibitemShut {NoStop}%
\bibitem [{\citenamefont {Dreyer}\ \emph {et~al.}(2004)\citenamefont {Dreyer},
  \citenamefont {Kelly}, \citenamefont {Krishnan}, \citenamefont {Finn},
  \citenamefont {Garrison},\ and\ \citenamefont
  {Lopez-Aleman}}]{Dreyer:2003bv}%
  \BibitemOpen
  \bibfield  {author} {\bibinfo {author} {\bibfnamefont {O.}~\bibnamefont
  {Dreyer}}, \bibinfo {author} {\bibfnamefont {B.~J.}\ \bibnamefont {Kelly}},
  \bibinfo {author} {\bibfnamefont {B.}~\bibnamefont {Krishnan}}, \bibinfo
  {author} {\bibfnamefont {L.~S.}\ \bibnamefont {Finn}}, \bibinfo {author}
  {\bibfnamefont {D.}~\bibnamefont {Garrison}}, \ and\ \bibinfo {author}
  {\bibfnamefont {R.}~\bibnamefont {Lopez-Aleman}},\ }\href {\doibase
  10.1088/0264-9381/21/4/003} {\bibfield  {journal} {\bibinfo  {journal}
  {Class. Quant. Grav.}\ }\textbf {\bibinfo {volume} {21}},\ \bibinfo {pages}
  {787} (\bibinfo {year} {2004})},\ \Eprint
  {http://arxiv.org/abs/gr-qc/0309007} {arXiv:gr-qc/0309007} \BibitemShut
  {NoStop}%
\bibitem [{\citenamefont {Berti}\ \emph {et~al.}(2009)\citenamefont {Berti},
  \citenamefont {Cardoso},\ and\ \citenamefont {Starinets}}]{berti_starinets}%
  \BibitemOpen
  \bibfield  {author} {\bibinfo {author} {\bibfnamefont {E.}~\bibnamefont
  {Berti}}, \bibinfo {author} {\bibfnamefont {V.}~\bibnamefont {Cardoso}}, \
  and\ \bibinfo {author} {\bibfnamefont {A.~O.}\ \bibnamefont {Starinets}},\
  }\href {\doibase 10.1088/0264-9381/26/16/163001} {\bibfield  {journal}
  {\bibinfo  {journal} {Class. Quant. Grav.}\ }\textbf {\bibinfo {volume}
  {26}},\ \bibinfo {pages} {163001} (\bibinfo {year} {2009})},\ \Eprint
  {http://arxiv.org/abs/0905.2975} {arXiv:0905.2975 [gr-qc]} \BibitemShut
  {NoStop}%
\bibitem [{\citenamefont {Abbott}\ and\ \citenamefont
  {et~al.}(2016)}]{PhysRevLett.116.221101}%
  \BibitemOpen
  \bibfield  {author} {\bibinfo {author} {\bibfnamefont {B.~P.}\ \bibnamefont
  {Abbott}}\ and\ \bibinfo {author} {\bibnamefont {et~al.}} (\bibinfo
  {collaboration} {LIGO Scientific and Virgo Collaborations}),\ }\href
  {\doibase 10.1103/PhysRevLett.116.221101} {\bibfield  {journal} {\bibinfo
  {journal} {Phys. Rev. Lett.}\ }\textbf {\bibinfo {volume} {116}},\ \bibinfo
  {pages} {221101} (\bibinfo {year} {2016})}\BibitemShut {NoStop}%
\bibitem [{\citenamefont {Abbott}\ \emph {et~al.}(2019)\citenamefont {Abbott}
  \emph {et~al.}}]{LIGOScientific:2019fpa}%
  \BibitemOpen
  \bibfield  {author} {\bibinfo {author} {\bibfnamefont {B.}~\bibnamefont
  {Abbott}} \emph {et~al.} (\bibinfo {collaboration} {LIGO Scientific,
  Virgo}),\ }\href {\doibase 10.1103/PhysRevD.100.104036} {\bibfield  {journal}
  {\bibinfo  {journal} {Phys. Rev. D}\ }\textbf {\bibinfo {volume} {100}},\
  \bibinfo {pages} {104036} (\bibinfo {year} {2019})},\ \Eprint
  {http://arxiv.org/abs/1903.04467} {arXiv:1903.04467 [gr-qc]} \BibitemShut
  {NoStop}%
\bibitem [{\citenamefont {Isi}\ \emph {et~al.}(2019)\citenamefont {Isi},
  \citenamefont {Giesler}, \citenamefont {Farr}, \citenamefont {Scheel},\ and\
  \citenamefont {Teukolsky}}]{Isi:2019aib}%
  \BibitemOpen
  \bibfield  {author} {\bibinfo {author} {\bibfnamefont {M.}~\bibnamefont
  {Isi}}, \bibinfo {author} {\bibfnamefont {M.}~\bibnamefont {Giesler}},
  \bibinfo {author} {\bibfnamefont {W.~M.}\ \bibnamefont {Farr}}, \bibinfo
  {author} {\bibfnamefont {M.~A.}\ \bibnamefont {Scheel}}, \ and\ \bibinfo
  {author} {\bibfnamefont {S.~A.}\ \bibnamefont {Teukolsky}},\ }\href {\doibase
  10.1103/PhysRevLett.123.111102} {\bibfield  {journal} {\bibinfo  {journal}
  {Phys. Rev. Lett.}\ }\textbf {\bibinfo {volume} {123}},\ \bibinfo {pages}
  {111102} (\bibinfo {year} {2019})},\ \Eprint
  {http://arxiv.org/abs/1905.00869} {arXiv:1905.00869 [gr-qc]} \BibitemShut
  {NoStop}%
\bibitem [{\citenamefont {Abbott}\ \emph {et~al.}(2020)\citenamefont {Abbott}
  \emph {et~al.}}]{Abbott:2020jks}%
  \BibitemOpen
  \bibfield  {author} {\bibinfo {author} {\bibfnamefont {R.}~\bibnamefont
  {Abbott}} \emph {et~al.} (\bibinfo {collaboration} {LIGO Scientific,
  Virgo}),\ }\href@noop {} {\  (\bibinfo {year} {2020})},\ \Eprint
  {http://arxiv.org/abs/2010.14529} {arXiv:2010.14529 [gr-qc]} \BibitemShut
  {NoStop}%
\bibitem [{\citenamefont {Berti}\ \emph {et~al.}(2016)\citenamefont {Berti},
  \citenamefont {Sesana}, \citenamefont {Barausse}, \citenamefont {Cardoso},\
  and\ \citenamefont {Belczynski}}]{Berti:2016lat}%
  \BibitemOpen
  \bibfield  {author} {\bibinfo {author} {\bibfnamefont {E.}~\bibnamefont
  {Berti}}, \bibinfo {author} {\bibfnamefont {A.}~\bibnamefont {Sesana}},
  \bibinfo {author} {\bibfnamefont {E.}~\bibnamefont {Barausse}}, \bibinfo
  {author} {\bibfnamefont {V.}~\bibnamefont {Cardoso}}, \ and\ \bibinfo
  {author} {\bibfnamefont {K.}~\bibnamefont {Belczynski}},\ }\href {\doibase
  10.1103/PhysRevLett.117.101102} {\bibfield  {journal} {\bibinfo  {journal}
  {Phys. Rev. Lett.}\ }\textbf {\bibinfo {volume} {117}},\ \bibinfo {pages}
  {101102} (\bibinfo {year} {2016})},\ \Eprint
  {http://arxiv.org/abs/1605.09286} {arXiv:1605.09286 [gr-qc]} \BibitemShut
  {NoStop}%
\bibitem [{\citenamefont {Inc.}()}]{Mathematica}%
  \BibitemOpen
  \bibfield  {author} {\bibinfo {author} {\bibfnamefont {W.~R.}\ \bibnamefont
  {Inc.}},\ }\href {https://www.wolfram.com/mathematica} {\enquote {\bibinfo
  {title} {Mathematica, {V}ersion 12.2},}\ }\BibitemShut {NoStop}%
\bibitem [{\citenamefont {Bhattacharyya}\ \emph {et~al.}(2016)\citenamefont
  {Bhattacharyya}, \citenamefont {Coates}, \citenamefont {Colombo},\ and\
  \citenamefont {Sotiriou}}]{Bhattacharyya:2015uxt}%
  \BibitemOpen
  \bibfield  {author} {\bibinfo {author} {\bibfnamefont {J.}~\bibnamefont
  {Bhattacharyya}}, \bibinfo {author} {\bibfnamefont {A.}~\bibnamefont
  {Coates}}, \bibinfo {author} {\bibfnamefont {M.}~\bibnamefont {Colombo}}, \
  and\ \bibinfo {author} {\bibfnamefont {T.~P.}\ \bibnamefont {Sotiriou}},\
  }\href {\doibase 10.1103/PhysRevD.93.064056} {\bibfield  {journal} {\bibinfo
  {journal} {Phys. Rev. D}\ }\textbf {\bibinfo {volume} {93}},\ \bibinfo
  {pages} {064056} (\bibinfo {year} {2016})},\ \Eprint
  {http://arxiv.org/abs/1512.04899} {arXiv:1512.04899 [gr-qc]} \BibitemShut
  {NoStop}%
\bibitem [{\citenamefont {Sago}\ \emph {et~al.}(2003)\citenamefont {Sago},
  \citenamefont {Nakano},\ and\ \citenamefont {Sasaki}}]{Sago:2002fe}%
  \BibitemOpen
  \bibfield  {author} {\bibinfo {author} {\bibfnamefont {N.}~\bibnamefont
  {Sago}}, \bibinfo {author} {\bibfnamefont {H.}~\bibnamefont {Nakano}}, \ and\
  \bibinfo {author} {\bibfnamefont {M.}~\bibnamefont {Sasaki}},\ }\href
  {\doibase 10.1103/PhysRevD.67.104017} {\bibfield  {journal} {\bibinfo
  {journal} {Phys. Rev. D}\ }\textbf {\bibinfo {volume} {67}},\ \bibinfo
  {pages} {104017} (\bibinfo {year} {2003})},\ \Eprint
  {http://arxiv.org/abs/gr-qc/0208060} {arXiv:gr-qc/0208060} \BibitemShut
  {NoStop}%
\end{thebibliography}%
\end{document}